\documentclass[a4paper,11pt]{article}
\usepackage{jcappub} 
\usepackage{lineno}
\makeatletter
\def\@fpheader{\relax}
\makeatother

\title{\boldmath Minimally coupled $\beta$-exponential inflation with an $R^2$ term in the Palatini formulation}

\author[a,b]{Nilay Bostan}
\author[c]{and Rafid H. Dejrah}
\affiliation[a]{Department of Physics and Astronomy, University of Iowa, 52242 Iowa City, IA, USA}
\affiliation[b]{Proton Accelerator Facility, Nuclear Energy Research Institute, \\ Turkish Energy Nuclear and Mineral Research Agency,  06980, Ankara, Türkiye}
\affiliation[c]{Department of Physics, Ankara University, Faculty of Sciences, 06100, Ankara, Türkiye}

\emailAdd{nilay.bostan@tenmak.gov.tr}
\emailAdd{rafid.dejrah@gmail.com}

\abstract{We focus on the inflationary predictions of $\beta$-exponential potential models, in which the inflaton is a representation of the field delineating the size of extra-dimension. Since it offers a well-motivated starting point for the study of physics at very high energies, we incorporate an $R^2$ term in the Palatini gravity. In addition, afterward the inflation, the inflaton oscillates about the minimum of the inflation potential, and reheats the universe. This occurs during the reheating phase, at which the inflaton decays into the standard model particles, which fill the universe. We extend our examination by considering the reheating effects on inflationary observables by employing the different scenarios of the reheat temperature. Supposing the standard thermal history after inflation, we display the inflationary predictions, $n_s, r, \mathrm{d}n_s/\mathrm{d}\ln k$ of $\beta$-exponential potential with minimal coupling in Palatini $R^2$ gravity. Also, different kinds of constraints from a variety of observations, such as BICEP/Keck, Planck 2018, as well as future possible detectable sensitivities that might be reached by CMB experiments: CMB-S4 and LiteBIRD are taken into account in this work. We indicate that our results are consistent with both the latest data and the future sensitivity forecasts of LiteBIRD/Planck and CMB-S4. Finally, the results in this study highlight the viability of our model even in the case of the existence of more stringent constraints expected from future achievable confidence level limits.}


\begin{document}
\maketitle
\flushbottom


\section{Introduction}
\label{introduction}
In the field of modern cosmology, there are a plethora of investigations for many concepts that assist us in gaining a deeper understanding of the universe. Amongst these nomenclatures comes ``inflation''; a period of exponential cosmic expansion of the universe's size and scale that occurred right after the big bang singularity. It is essential to advance our studies of the notion of inflation because the scientific community has focused on comprehending and characterizing some properties of the origin of the universe for many years. An excellent comprehension of the inflationary epoch is necessary to accomplish this. For the sake of example, the concept is taken into account in ref.~\cite{Linde:2007fr} with details. Furthermore, the idea of inflation is very pivotal to explicate a legion of what were thought to be issues for cosmologists, for instance; the structure problem, smoothness problem, flatness problem, large-scale structure problem, and other issues. It is worth mentioning that the small-scale structure problem still exists, and is not solved by inflation. For more details on how the cosmic concept of inflation participated in solving such issues, see the following studies~\cite{Starobinsky:1979ty, Linde:1983gd}. Hence, inflation is considered by the majority to be a milestone for modern cosmology, especially for its merit that it can be extended to explain a wide range of other concepts; see ref.~\cite{Mukhanov:1981xt} for more detailed cases. Not only does it achieve that, but it also advances it to a more acceptable status by being supported by measurements from the cosmic microwave background (CMB) anisotropies, notwithstanding the ongoing discussion about related frameworks. 

One can relate to the fact that exponential models are used to depict cosmic inflation since it is explained as the universe expanding quasi-exponentially in the early era. An exponential model is a mathematical description in which the universe's scale factor increases exponentially with time, and it is employed to represent the rapid expansion of the universe during the inflationary period, where the scale factor $a(t)$ grows approximately as $a(t) \propto e^{Ht}$, and $H$ is being the Hubble parameter. This exponential expansion aids in elucidating the observed large-scale homogeneity and isotropy of the universe, see the refs.~\cite{Guth:1980zm, Starobinsky:1980te, Linde:2007fr}. In the literature, a variety of distinct models with different special properties, potentials, and inflation fields are taken into consideration in very detailed procedures; see refs.~\cite{Martin:2013tda, Benetti:2016jhf, SantosdaCosta:2017ctv, Benetti:2019kgw, SantosdaCosta:2020dyl}. In this work, we consider the $R^2$ term in the Einstein-Hilbert action, which was first introduced by A. Starobinsky (ref.~\cite{Starobinsky:1980te}). The $R^2$ term gained acceptance because it can be derived for the inflationary expansion by introducing a scalar degree of freedom known as the scalaron, which acts a role of the inflaton; hence, the model does not require a separate, ad-hoc inflation field to achieve inflation, making it one of the earliest and most accomplished inflationary models. Moreover, amongst the variations and principles, (it can be dissected in the refs.~\cite{Misner1973, Wald:1984rg}) that one can apply to the Einstein-Hilbert action to derive Einstein's equations, we are going to work in this paper with the Palatini formalism [on the contrary to the name, it was Einstein who introduced it, see ref.~\cite{Ferraris:1982wci}]. Palatini formalism is defined as an independent variation with respect to the metric, an independent connection, and reduced standard deviation. It is remarkable to mention that theories based on this formalism satisfy the metric postulates \cite{Will:2018bme}. The importance of the Palatini formalism is that it has been demonstrated to supply intriguing phenomenological implications. As one of the most pivotal examples, the differences between the metric and the Palatini formulations for inflationary predictions, can be given~\cite{Bauer:2008zj, Racioppi:2019jsp}. According to studies in the literature, the Palatini formalism predicts the inflationary observables, especially for the tensor-to-scalar ratio ($r$) which makes it more prevailing than the metric formulation~\cite{Bauer:2008zj,Bostan:2020pnb}. Also, see refs.~\cite{Jarv:2017azx, Bostan:2019uvv, Racioppi:2019jsp, Bostan:2022swq, Bostan:2023ped, Dioguardi:2022oqu, Ghoshal:2024ycp} that delve into the inflation in Palatini formalism with details. These features can potentially offer better alignment with the measurements from CMB anisotropies and large-scale structure surveys, see the refs.~\cite{Koivisto:2005yc, Borunda:2008kf, Gialamas:2023flv}. Thus, by considering this type of formalism principle, our work can leverage the mentioned advantages and, in return, provide us with a robust and comprehensive analysis of the inflationary dynamics in the context in which our work is set.

In the literature, $\beta$-exponential inflation has already been considered in many studies so far. For this point of view, the pivotal study is the ref.~\cite{Alcaniz:2006nu}, which investigates the inflationary predictions of $\beta$-exponential potential in a minimally coupled case. They depict the trajectories for disparate values of $\beta$ parameters in the $n_s-r$ and $\mathrm{d}n_s/\mathrm{d}\ln k-n_s$ planes for some chosen values of $\beta$ and compare their findings with the cosmological data. In addition to ref.~\cite{Alcaniz:2006nu}, there are two important studies in the literature that inspect the $\beta$-exponential inflation in detail; see the following refs.~\cite{Santos:2017alg, dosSantos:2021vis}. Ref.~\cite{Santos:2017alg} shows the $n_s-r$ plane for different values of $\beta$, as well as some selected values of $\lambda$, considering two different number of e-folds $N_*$: $50$ and $60$, and comparing the predictions of this model with the Planck data. Additionally, ref.~\cite{dosSantos:2021vis} has calculated the inflationary predictions for the non-minimally coupled $\beta$-exponential inflation and compared their results with the CMB data. They present the cosmological consequences of the non-minimally coupled $\beta$-exponential inflation with details in the metric formulation. [Recent studies that also include the $\beta$-exponential potential model are as follows: ~\cite{Santos:2022exm, Capistrano:2024kuc}]. The potential model of $\beta$-exponential inflation is very pivotal to take into account because the model can appear in the framework of brane cosmology at which the inflaton, see ref.~\cite{Martin:2013tda} for the models that have been examined so far based on inflaton, is regarded as the field representing the size of extra-dimension. It is suitable to mention here that this type of potential model is derived using the braneworld scenario framework. For more details about the concept of inflation in brane cosmology, see the following refs.~\cite{Dvali:1998pa, Santos:2017alg}. Here, we refer to some insightful studies for a detailed analysis, regarding: \\
i) Examining the Inflating Branes Concept, see ref.~\cite{Mersini-Houghton:2000pof}. \\
ii) Phenomenological potentials in the 3-brane scenario, see the refs.~\cite{Randall:1999ee, Goldberger:1999uk}. \\ On the other hand, the warm inflation scenario is investigated by the class of $\beta$-exponential potentials with details in ref.~\cite{Santos:2022exm}.

Furthermore, the couplings between the inflaton and the standard model (SM) particles are essential in indicating the dynamics of the reheating phase; a phase that transitions the universe from inflationary expansion to a hot, radiation-dominated era; see the following refs.~\cite{Abbott:1982hn, Albrecht:1982mp, Kofman:1997yn, Chung:1998rq, Bassett:2005xm, Dai:2014jja, Munoz:2014eqa, Cook:2015vqa, Hanin:2023ypf} for more details about the notion itself. In addition, these couplings result in the production of SM particles leading to the impact on the thermalization process and the subsequent evolution of the universe. The inflaton couples to other fields throughout the reheating phase, converting the remaining energy into new particles that make up the radiation energy density, see refs.~\cite{Dolgov:1989us, Traschen:1990sw, Kofman:1994rk, Saha:2020bis}. Across this manuscript, we thoroughly analyze the reheating effects on the inflationary predictions within the context of our model in order to provide a comprehensive understanding of the reheating dynamics by calculating the inflationary observables for different reheat temperatures. [For details about the reheating dynamics, see the following refs.~\cite{Kofman:1994rk, Allahverdi:2010xz, Amin:2014eta}]. Moreover, in this work, we depict inflationary predictions with reheating impacts that can have a consistency of our model with current observational data. In literature, refs.~\cite{Drewes:2015coa, Repond:2016sol, Aoki:2022dzd} can be examined for more details about the interactions and their effects between different models and the inflaton. Additionally, these references~\cite{Kallosh:1999jj, Bassett:2005xm, PhysRevD.73.023501} provide the results for further analysis related to the reheating process and the subsequent cosmological observables.

In this manuscript, we study the inflationary predictions for the potential model that generalizes the well-known power law inflation, see the refs.~\cite{Abbott:1984fp, Lucchin:1984yf, Ratra:1987rm, Ferreira:1997hj, Martin:2013tda}, through a general exponential function ~\cite{Alcaniz:2006nu, Martin:2013tda}, which is the $\beta$-exponential potential. The framework of braneworld scenarios can be used to generate this potential, which can be accurately compared with the observational data \cite{Alcaniz:2006nu, Santos:2017alg, dosSantos:2021vis}. In this work, we specifically study this potential in the minimally coupled case with an $R^2$ term in Palatini formalism \cite{Enckell:2018hmo}. The specific case of the Palatini formalism lies in its ability to significantly reduce the tensor-to-scalar ratio $r$ in regions of the parameter space~\cite{Bostan:2019uvv}. Unlike the metric formulation which predicts large values of the tensor-to-scalar ratio $r$, which are not favored by observational data. The feature of the Palatini formulation makes it a privileged choice for inflationary models since it aligns better with current measurements from the CMB anisotropies. When comparing these two variation methods, one can see how the inflationary models are more viable with the Palatini formalism for the merits mentioned above. For more results that support this point, please see tables 2 and 3 in refs.~\cite{, Aoki:2024jha, Ghoshal:2024ycp} which studied this case in detail. Additionally, one of the motivations to work with Palatini formulation is that the symmetries of the fundamental action are more manifest in this type of approach as discussed in ref.~\cite{Anselmi:2020opi}. For this reason, we can get a more insightful framework for constructing and analyzing the inflationary models.

We analyze the inflationary observables of this potential and compare our results for the inflationary predictions with the current data from Planck and BICEP/Keck \cite{BICEP:2021xfz}, as well as the future CMB-S4 \cite{Abazajian:2019eic} and LiteBIRD \cite{LiteBIRD:2022cnt} sensitivity forecasts. On the other hand, it is good to mention here ref.~\cite{Antoniadis:2018yfq}, which recently indicates that the fundamental theory of gravity is a Palatini as opposed to being a metric when the gravity sector is extended by an $\alpha R^2$ term (where $\alpha$ is a dimensionless parameter).  It is worth mentioning that the inclusion of the $\alpha R^2$ term is not primarily motivated by renormalizability, but rather by its ability to drive inflation and improve the ultraviolet (UV) behavior of the theory. Therefore, adding an $\alpha R^2$ term assures a well-motivated starting point for the physics analysis at very high energies \cite{Tenkanen:2019jiq}. In literature, the studies that discuss the Palatini $R^2$ inflation can be listed as follows: \cite{Karam:2018mft, Antoniadis:2018ywb, Antoniadis:2019jnz, Tenkanen:2019jiq, Gialamas:2020snr, Karam:2021sno, Dimopoulos:2022rdp, Ghoshal:2022qxk, Aoki:2024jha}. In addition, ref.~\cite{Ghoshal:2024ycp} has considered the post-inflationary leptogenesis and the production of dark matter in the Palatini formalism with all details.

The paper is mapped as follows. In section~\ref{section2}, we introduce the framework we are going to work on, such as introducing the Einstein-Hilbert action that includes the $R^2$ term in the Palatini formalism for different frames. Moreover, we introduce the $\beta$-exponential inflation model, and its mathematical properties to provide a better understanding for our analytical and numerical analysis later on. We also introduce the Einstein frame potential form in the section alongside an illustrated study for a variety of choices of $\beta$ to get a better image. The slow-roll parameters are provided in both the canonical scalar field $\zeta$ and in the terms of the original scalar field $\phi$. The number of e-folds $N_*$ is also given in two different forms for both analytical and numerical calculations, and to get better results for the latter one the reheat temperature concept $T_{reh}$ is also introduced, for three different scenarios. We have also introduced the brane inflation with detailed analysis in section~\ref{section2}, since it is connected to the context of our paper. In section~\ref{results}, we show and discuss our analytical and numerical results by illustrating a thorough figure that helps visualize our analytical results and then a detailed table by which one can get a clearer image of our discussions. We have considered a variety of observational data choices as well. Finally, we summarize and conclude the paper in section~\ref{conc}. In the appendix section~\ref{sec: appendix}, we provide related analytical approximations of section \ref{results}. We adopt $M_{\rm P}$ to unity for our calculations\footnote{In our analysis, we utilize natural units where $M_{\rm P} \equiv 1$.} that we depict with details in section~\ref{results}.

\section{Palatini $\beta$-exponential inflation with an $R^2$ term}
\label{section2}
We commence by acquainting the action that is taken into consideration in this work~\cite{Antoniadis:2018yfq, Tenkanen:2019jiq}

\begin{equation}
    \label{nonminimal_action1}
	S_J = \int d^4x \sqrt{-g}\left(\frac{1}{2}M_{\rm P}^2R + \frac{\alpha}{4} R^2 - \frac{1}{2} \nabla^{\mu}\phi\nabla_{\mu}\phi - V(\phi) \right) \,,
\end{equation}

where $g$ is the determinant of the metric tensor $g_{\mu\nu}$, $J$ indicates that the action is given in the Jordan frame. $R$ is the Ricci scalar, which is defined by $R = g^{\mu\nu} R_{\mu\nu} (\Gamma)$, where $R_{\mu\nu}$ is the Ricci tensor derived by using the Christoffel symbols, $\Gamma_{\mu\nu}^\lambda$. Also, $\phi$ is the scalar field, called the inflaton, and $V(\phi)$ is the potential given in the Jordan frame. $M_P$ is the reduced Planck mass, and $\alpha$ is the dimensionless parameter. The Jordan frame action given in eq. (\ref{nonminimal_action1}) can be defined in terms of the auxiliary scalar field $\chi$ dynamically, as follows~\cite{Antoniadis:2018yfq, Tenkanen:2019jiq}:
\begin{eqnarray}
    S_J &=& \int d^4x \sqrt{-g}\bigg (\frac{1}{2}M_{\rm P}^2\left(1+\alpha \chi^2 \right) R -\frac{\alpha}{4}\chi^4 - \frac{1}{2}\nabla^{\mu}\phi\nabla_{\mu}\phi - V(\phi) \bigg )\,.
\end{eqnarray}
In addition, one can switch from the Jordan frame ($J$) to the Einstein ($E$) frame by applying a Weyl rescaling~\cite{Fujii:2003pa}. By performing the Weyl transformation of the metric with
\begin{equation}
    \label{Omega1}
	\Tilde{g}_{\mu\nu} \to \Omega g_{\mu\nu}, \hspace{1cm} \Omega \equiv 1+\frac{\alpha \chi^2}{M_{\rm P}^2}\,,
\end{equation}

one can write the action within the Einstein frame, resulting in \cite{Antoniadis:2018yfq}:
\begin{equation}
\label{einsteinframe1}
	S_E = \int d^4x \sqrt{-\Tilde{g}}\left(\frac{1}{2}M_{\rm P}^2 \Tilde{R} - \frac{1}{2\Omega}\nabla^{\mu}\phi\nabla_{\mu}\phi - \left(\frac{M_{\rm P} }{\Omega}\right)^2 V(\phi, \chi) \right) \,,
\end{equation} 

here $V(\phi, \chi) = V(\phi) + \frac{\alpha}{4}\chi^4 $. Variation of this action given in eq. \eqref{einsteinframe1} with respect to the $\chi$ is obtained for the constraint equation~\cite{Antoniadis:2018ywb, Antoniadis:2019jnz, Tenkanen:2019jiq, Antoniadis:2018yfq}
\begin{equation}
\label{auxiliary}
\frac{\delta S_E}{\delta \chi} = 0 \ \rightarrow \  \frac{\chi^2}{M_{\rm P}^2} = \frac{4V(\phi) + \nabla^{\mu}\phi\nabla_{\mu}\phi}{M_{\rm P}^4 - \alpha \nabla^{\mu}\phi\nabla_{\mu}\phi}.
\end{equation}

Substituting eq. \eqref{auxiliary} into eq. \eqref{einsteinframe1}, one can obtain the form \cite{Enckell:2018hmo}: 
\begin{equation}
\label{einsteinframe2}
	S_E \simeq \int d^4x \sqrt{-\Tilde{g}}\left(\frac{1}{2}M_{\rm P}^2 \Tilde{R} - \frac{1}{2}\frac{\nabla^{\mu}\phi\nabla_{\mu}\phi}{\left(1 +  \frac{4\alpha}{M_{\rm P}^4}V(\phi)\right)} - \frac{V(\phi)}{\left(1 +  \frac{4\alpha}{M_{\rm P}^4}V(\phi)\right)} \right) \,.
\end{equation}

It is important to mention here, by making the following field redefinition,
\begin{equation} \label{scalardefinit}
\mathrm{d}\zeta = \frac{\mathrm{d}\phi}{\sqrt{1 + \frac{4\alpha}{M_{\rm P}^4}V(\phi)}} = \frac{\mathrm{d}\phi}{\sqrt{Z(\phi)}},
\end{equation}

the action for a minimally coupled scalar field $\zeta$ with a canonical kinetic term can be obtained. Here, $Z(\phi) = 1 + \frac{4\alpha}{M_P^4} V(\phi)$ is known as the field space metric. Eq. \eqref{scalardefinit} can be computed with respect to the form of the specific potential models that are considered. In addition, we can indicate the Einstein frame potential from the action described in eq. \eqref{einsteinframe2} as follows:
\begin{equation} \label{eframebeta}
    V_E(\phi) = \frac{V(\phi)}{\left(1 + \frac{4\alpha}{M_{\rm P}^4}V(\phi)\right)}.
\end{equation}

As one can notice from eq. \eqref{eframebeta}, the potential $V(\phi)$ is scaled by a factor as written in the denominator of the equation, which is a consequence of the existence of the term $R^2$.

In this work, we focus on the inflationary predictions of the $\beta$-exponential potential, which we describe in the following section. We display the inflationary parameters for this potential, the spectral index $n_s$, the tensor-to-scalar ratio $r$, and the running of the spectral index $\mathrm{d}n_s/\mathrm{d}\ln k$, by supposing the standard thermal history afterward inflation, and for this potential, we present the compatible regions for the spectral index $n_s$ and the tensor-to-scalar ratio $r$ within the recent Planck + BICEP/Keck data and the future CMB-S4 and LiteBIRD/Planck achievable sensitivity forecasts. We show the cosmological consequences of $\beta$-exponential inflation with an $R^2$ term in Palatini formalism, presenting the results of inflationary predictions of this potential.
\subsection{$\beta$-exponential inflation}
In this work, we study the $\beta$-exponential potential model, which was first introduced and studied in ref.~\cite{Alcaniz:2006nu} as a generalization of the power law inflation phenomenological~\cite{Abbott:1984fp, Ratra:1987rm, Ferreira:1997hj, Martin:2013tda}, which is already mentioned in the introduction. We start constructing our model with the usual exponential function \cite{Martin:2013tda} which is defined by
\begin{equation} \label{potexp}
V(\phi) = M^4 \exp{(-\lambda \phi / M_{\rm P})}\; .
\end{equation}

\emph{In this work}, we discuss a possible generalization for the inflation potential, which is given in eq. (\ref{potexp}), with the following form:
\begin{equation} \label{potexpg}
V(\phi) = M^4 \exp_{1-\beta}{(-\lambda \phi / M_{\rm P})}\; ,
\end{equation}
where the definition of the generalized exponential function $ \exp_{1-\beta}$ is as follows~\cite{abramowitz1968handbook,  Lima:2001lgd, Alcaniz:2006nu}

\begin{equation} \label{potexpgdef}
\exp_{1-\beta}{({f})} = \left[1 + \beta {f} \right]^{1/\beta}\; ,
\end{equation}
\begin{eqnarray}
\mbox{for} \; \left\{
\begin{tabular}{l}
$1 + \beta {{f}} > 0$\\
\\
$\exp_{1-\beta}({f}) = 0$, otherwise.
\end{tabular}
\right.
\nonumber
\end{eqnarray}
For $f > 0$ and $g > 0$, this function satisfies the following identities (as it is already delineated and discussed with details in~\cite{Alcaniz:2006nu, Martin:2013tda}):
\begin{equation}
\label{p1}
\exp_{1-\beta}\left[\ln_{1-\beta}({f}) \right] =  {f}\;, \nonumber
\end{equation}
and
\begin{equation}
\label{p2}
\ln_{1-\beta}({f}) + \ln_{1-\beta}({g}) =   \ln_{1-\beta}({fg}) -  \beta \left[\ln_{1-\beta}({f}) \ln_{1-\beta}({g})\right], \nonumber
\end{equation}
where $\ln_{1-\beta}({f}) = (f^{\beta} - 1)/\beta$ is the generalized logarithmic function. The $\beta$-exponential potential can fulfill the disruption of the slow-roll regime with the end of inflation~\cite{Santos:2022exm}, thus it makes the tiny values for the tensor-to-scalar ratio, $r$ (\cite{Alcaniz:2006nu, Santos:2022exm}). 

The Jordan frame potential, $V(\phi)$, for the $\beta$-exponential inflation can be written in the following form\footnote{In the framework of brane cosmology, where the radion (a field characterizing the size of the extra-dimension) is elucidated as the inflaton, the $\beta$-exponential potential in eq. \eqref{jframebetamodel} can arise, for more details see ref.~\cite{Santos:2017alg}. The important point to emphasize is that the brane inflation scalar potential can be easily obtained from the effective potential which is described by the four-dimensional action as follows \cite{Santos:2017alg}:
\begin{equation*}\label{action-4d}
S_4 = \int{d^4x}\sqrt{-g_4}\left(\frac12\sigma\dot{L}^2-V_{\rm eff}(L)\right)\;,
\end{equation*}
where $\sigma$ is the brane tension, $L$ is the position of the brane with respect to $r = 0$, here $r$ is the fifth coordinate of the 5D bulk. Also, the effective potential is defined as $V_{\rm eff}(L)=V_0(1+c_1L)^{\frac{1}{\lambda c_1}}+\frac12\sigma$. In addition, with some changes in the definitions of the effective potential, $V_{\rm eff}(L)$, the equivalent form of the $\beta$-exponential potential, which is presented in eq. \eqref{jframebetamodel} can be found, see ref.~\cite{Santos:2017alg}. According to the studies refs.~\cite{Santos:2017alg,dosSantos:2021vis}, it can be inferred that the brane tension $\sigma$ is related to the ratio $\beta / \lambda$. Both $\beta$ and $\lambda$ are constrained by this connection; that is, $\beta$ must be greater than $\lambda$, with $\beta \geq 1/2$. Thus, considering the $\beta$-exponential potential models is highly motivated for the context of braneworld scenarios and brane dynamics~\cite{dosSantos:2021vis}. For more details thoroughly, please see ~\cite{Santos:2017alg,dosSantos:2021vis}.}, where the constraints provided by eq. \eqref{potexpg} are taken into consideration~\cite{Lillepalu:2022knx}:
\begin{equation} \label{jframebetamodel}
    V(\phi) = V_0 \left(1-\lambda \beta \frac{\phi}{M_{\rm P}}\right)^{1/\beta},
\end{equation}
where the deviation from the pure exponential function is controlled by constant $\beta$, while $\lambda$ is a dimensionless constant. 

\begin{figure}[t!]
	\centering
	\includegraphics[angle=0, width=13.0cm]{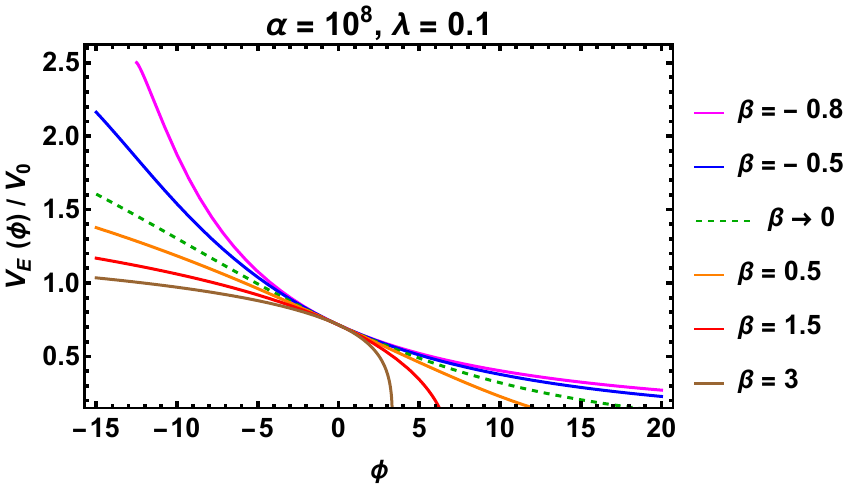}
	\caption{The Einstein frame $\beta$-exponential potential with minimal coupling in Palatini $R^2$ gravity as a function of $\phi$. The colors show different values of the $\beta$ parameter. We fixed $\lambda = 0.1$, $\alpha = 10^{8}$, $V_0=10^{-9}$, as well as taken $M_{\rm P}$ unity. 
}
	\label{fig:Eframepot}
\end{figure}

In this work, with the form of eq. (\ref{eframebeta}), we delve into the Einstein frame minimally coupled potential for the $\beta$-exponential inflation with an $R^2$ term in the Palatini formalism. By using eq. \eqref{jframebetamodel}, we can define the potential model in the Einstein frame as follows:
\begin{equation} \label{eframebetamodel}
        V_E(\phi) = \frac{V_0 \left(1-\lambda \beta \frac{\phi}{M_{\rm P}}\right)^{1/\beta}}{\left(1 + 4\alpha \frac{V_0 \left(1-\lambda \beta \frac{\phi}{M_{\rm P}}\right)^{1/\beta}}{M_{\rm P}^4}\right)}.
\end{equation}
In figure~\ref{fig:Eframepot}, we illustrate how the Einstein frame $\beta$-exponential potential, which is given in eq. (\ref{eframebetamodel}) changes according to the values of the $\beta$ parameter, which we select and how this parameter controls the potential model, as well as deviation from the usual exponential function. 

\subsection{Inflationary observables} \label{inflationary_observables}
As long as the Einstein frame potential can be obtained in terms of the canonical scalar field $\zeta$, inflationary predictions can be calculated by exploiting the slow-roll parameters as follows~\cite{Lyth:2009zz}:
\begin{equation}\label{slowroll1}
\epsilon =\frac{M_{\rm P}^2}{2}\left( \frac{V_{\zeta} }{V}\right) ^{2}\,, \quad
\eta = M_{\rm P}^2 \frac{V_{\zeta \zeta} }{V}  \,, \quad
\kappa ^{2} = M_{\rm P}^4 \frac{V_{\zeta} V_{\zeta\zeta\zeta} }{V^{2}}\,,
\end{equation}
where $\zeta$ in the subscript denotes the derivatives with respect to the canonical scalar field $\zeta$. Here, $\epsilon$ is the parameter that measures the steepness of the inflationary potential, and it is an indicator of how quickly the scalar field $\zeta$ is rolling down potential. The lower values of $\epsilon$ results in slower roll and more sustained inflation. $\eta$, on the other hand, is the parameter that describes the slope of the potential changes that affect the stability and duration of inflation. Moreover, $\kappa^2$ investigates finer details about the shape of potential and the dynamics of the inflation, in instances refs.~\cite{Casadio:2005xv, Forconi:2021que, Lambiase:2023ryq} have studied these parameters in detail. Inflationary observables, for example, the spectral index $n_s$, the tensor-to-scalar ratio $r$, and the running of the spectral index $\mathrm{d} n_s/\mathrm{d} \ln k$ are given in the following forms for the slow-roll approximation:
\begin{eqnarray}\label{nsralpha1}
n_s = 1 - 6 \epsilon + 2 \eta \,,\quad
r = 16 \epsilon, \nonumber\\
\frac{\mathrm{d}n_s}{\mathrm{d}\ln k} = 16 \epsilon \eta - 24 \epsilon^2 - 2 \kappa^2\,.
\end{eqnarray}
Recently, more precise constraints on the inflationary predictions have been provided by BICEP/Keck~\cite{BICEP:2021xfz}, especially for the tensor-to-scalar ratio $r$, which tightens to $r < 0.035$ at $95\%$ CL. This strong constraint elucidates the amplitude of the primordial gravitational waves, as well as the inflationary scale. Moreover, recent BICEP/Keck results also constrain the spectral index $n_s$ to the range $[0.957, 0.976]$ at $2\sigma$ of confidence level. These constraints are with the pivot scale, which is selected at $k_* = 0.002$ Mpc$^{-1}$. In addition, the next generation of CMB surveys, such as CMB-S4~\cite{Abazajian:2019eic}, which aims for $r \simeq \mathcal{O}(10^{-3})$. Also, the future measurement from the LiteBIRD experiment~\cite{LiteBIRD:2022cnt} will be able to test the inflationary models precisely. Moreover, another pivotal constraint arises from the Planck 2018 measurements along with the results from the baryon acoustic oscillations (BAO). They provide the constraint on $\mathrm{d}n_s / \mathrm{d}\ln k = -0.0041 \pm 0.0067$ to base $\Lambda$CDM in $68\%$, TT,TE,EE +lowE+lensing+BAO~\cite{Planck:2018vyg}. In the future, some improvements are expected from the observations of the 21-cm line, ~\cite{Kohri:2013mxa, Basse:2014qqa, Munoz:2016owz}. In addition, for the case of the tensor-to-scalar ratio $r > 0.003$, at larger than $5\sigma$ of confidence level, primordial gravitational waves can be detectable in the future by CMB-S4~\cite{Abazajian:2019eic}. The highest limit of the tensor-to-scalar ratio $r < 0.001$ at $95\%$ CL may be reached through future observations done by CMB-S4, even in the absence of a detection, this limit would still provide important new insights for the inflation~\cite{Abazajian:2019eic}. 

The number of e-folds $N_*$ in the slow-roll approximation is given by
\begin{equation} \label{efold1}
N_*=\frac{1}{M_{\rm P}^2}\int^{\zeta_*}_{\zeta_e}\frac{V\rm{d}\zeta}{V_{\zeta}}\,, \end{equation}
where the subscript ``$_*$'' denotes the quantities when the pivot scale exits the horizon, and $\zeta_e$ is the value of the inflaton at which the inflation ends, we can compute $\zeta_e$ via $\epsilon(\zeta_e) =1$. 

Furthermore, the amplitude of the curvature perturbation can be calculated by using the following relation:
\begin{equation} \label{perturb1}
\Delta_\mathcal{R}=\frac{1}{2\sqrt{3}\pi M_{\rm P}^3}\frac{V^{3/2}}{|V_{\zeta}|},
\end{equation}
the best fit value for the pivot scale $k_* = 0.002$ Mpc$^{-1}$ is $\Delta_\mathcal{R}^2\approx 2.1\times10^{-9}$~\cite{Planck:2018vyg}, which is obtained from the Planck measurements.

On the other hand, it may not always be easy or possible to obtain the analytical expression of an inflation potential defined as $V_J(\phi)$ in the Jordan frame as $V_E(\zeta)$ in the Einstein frame. This depends on the form of the inflation potential which is taken into account. In this case, the analysis of its predictions for the considered potential can be made in terms of the original scalar field $\phi$ numerically instead of the canonical scalar field $\zeta$. Further analysis related to $\zeta$ is mentioned in detail in section~\ref{results}. We use such equations to perform numerical calculations in terms of the scalar field $\phi$ when necessary. Furthermore, for the numerical calculations, one needs to have the slow-roll parameters in terms of the field $\phi$ to be able to calculate the inflationary predictions of the potential model in terms of the general values of free parameters. Thus, the slow-roll parameters should be acquired in terms of the scalar field $\phi$, and these parameters can be written as follows~\cite{Linde:2011nh}:
\begin{eqnarray}\label{slowroll2}  
\epsilon=Z\epsilon_{\phi}\,, \ \
\eta=Z\eta_{\phi}+{\rm sgn}(V')Z'\sqrt{\frac{\epsilon_{\phi}}{2}}\,,\nonumber\\
\kappa^2=Z\left(Z\kappa^2_{\phi}+3{\rm sgn}(V')Z'\eta_{\phi}\sqrt{\frac{\epsilon_{\phi}}{2}}+Z''\epsilon_{\phi}\right),
\end{eqnarray}
where the slow-roll parameters are defined in terms of $\phi$ as the following:
\begin{equation}
\epsilon_{\phi} =\frac{1}{2}\left( \frac{V^{\prime} }{V}\right) ^{2}\,, \quad
\eta_{\phi} = \frac{V^{\prime \prime} }{V}  \,, \quad
\kappa ^{2} _{\phi}= \frac{V^{\prime} V^{\prime \prime\prime} }{V^{2}}\,.
\end{equation}
Here `` $^\prime$ '' represents the derivatives with respect to the original scalar field $\phi$. Furthermore, eqs. \eqref{efold1} and \eqref{perturb1} can be written with regard to $\phi$ resulting in the following forms:
\begin{eqnarray}\label{perturb2}
N_*&=&\rm{sgn}(V')\int^{\phi_*}_{\phi_e}\frac{\mathrm{d}\phi}{Z(\phi)\sqrt{2\epsilon_{\phi}}}\,,\\
\label{efold2} \Delta_\mathcal{R}&=&\frac{1}{2\sqrt{3}\pi}\frac{V^{3/2}}{\sqrt{Z}|V^{\prime}|}\,.
\end{eqnarray}
To calculate the numerical values of observables; the spectral index $n_s$, the tensor-to-scalar ratio $r$, and the running of the spectral index $\mathrm{d}n_s / \mathrm{d}\ln k$, the numerical value of the number of e-folds $N_*$ is also required. Supposing a standard thermal history afterward inflation, one can express the number of e-folds $N_*$ for the pivot scale $k_* = 0.002$ Mpc$^{-1}$ in the following form~\cite{Liddle:2003as}:
\begin{eqnarray} \label{efolds}
N_*\approx64.7+\frac12\ln\frac{\rho_*}{M_{\rm P}^4}-\frac{1}{3(1+\omega_r)}\ln\frac{\rho_e}{M_{\rm P}^4} +\left(\frac{1}{3(1+\omega_r)}-\frac14\right)\ln\frac{\rho_r}{M_{\rm P}^4},
\end{eqnarray}
here, $\rho_{e}=(3/2)V(\phi_{e})$ is the energy density at the end of inflation, $\rho_r$ is the energy density at the end of reheating, and $\rho_*\approx V(\phi_*)$ is the energy density when the scale corresponding to $k_*$ exits the horizon. $\omega_r$ is the equation of the state parameter during reheating. The definitions of $\rho_{r}$ and $\rho_*$ are given in the following forms:
\begin{equation}
 \rho_{r} = \Big(\frac{\pi^2}{30}g_*\Big) T_{reh}^4,  \qquad  \rho_{*} = \frac{3 \pi^2\Delta^2_\mathcal{R} r }{2},
\end{equation}
where the standard model value $g_*=106.75$, which gives the number of relativistic degrees of freedom, can be employed to compute $\rho_{r}$. Also, $T_{reh}$ indicates the reheat temperature; the temperature at which the universe is in thermal equilibrium and radiation dominates. At the end of the reheating phase, thermal equilibrium is reached and the universe is fully filled with radiation~\cite{Baumann:2022mni}. The inflation potential represents the majority of the universe's energy density during inflation. When the potential steepens and the inflation field (inflaton) gains kinetic energy, inflation terminates. The SM particles then need to receive the energy from the inflaton sector. In addition, the hot big bang is initiated by reheating. The inflaton's decay produce a particle soup that eventually approach thermal equilibrium, with the radiation and particle fields present at the time, ensuring that energy is uniformly distributed across the universe's constituents~\cite{Mukhanov:2005sc}, at a certain temperature, \textit{reheat temperature,} as a result of particle interactions. The energy density $\rho_r$ at the end of the reheating period gives this reheat temperature. For much more details and highlights on the thermal history of the universe and reheating, please see the following refs.~\cite{Baumann:2009ds, baumann2014cosmology, Baumann:2022mni}. Moreover, here are the detailed studies related to the reheating concept and constraints on inflationary predictions \cite{Bostan:2018evz, Bostan:2024fyz}. 

In this work, we consider two different cases for the number of e-folds $N_*$ which is defined in eq. \eqref{efolds}:
\begin{itemize}
    \item \underline{First case:} We take $w_r=1/3$, it is the instant reheating assumption. With this assumption, the number of e-folds $N_*$ in  eq. \eqref{efolds} reduces to the following form:
\begin{eqnarray} \label{efoldsreal2} 
N_*\approx64.7+\frac12\ln\frac{\rho_*}{M_{\rm P}^4}-\frac14\ln\frac{\rho_e}{M_{\rm P}^4}.
\end{eqnarray}

Equation (\ref{efoldsreal2}) demonstrates that the inflationary predictions should not depend on $T_{reh}$ for the instant reheating assumption.
    \item \underline{Second case:} We take $w_r=0$, and with this selection, the number of e-folds $N_*$ should depend on the reheat temperature since the potential does not have a minimum, then the reheating cannot happen in a standard way. In our numerical calculations, we take $T_{reh} = 10^{8}$ GeV and $T_{reh} =10^{14}$ GeV in the following section. Ref.~\cite{Dimopoulos:2022rdp} has studied the reheating mechanisms for the quintessential inflation in Palatini $R^2$ gravity, they have found the maximum reheating temperature to be $T_{reh} =10^{14}$ GeV, accordingly in this study we select $10^{14}$ GeV as the highest reheat temperature, as well as we show the predictions for the $T_{reh} = 10^8$ GeV because, at this specific limit, our model works and consistent with the cosmological data in a good manner. It is also important to note that here, in the second case, the value of the e-folds number is less than the ones for the case when $w_r=1/3$ (instant reheating). By taking $w_r=0$, eq. \eqref{efolds} becomes:
    \begin{eqnarray} \label{efolds3}
N_*\approx64.7+\frac12\ln\frac{\rho_*}{M_{\rm P}^4}-\frac{1}{3}\ln\frac{\rho_e}{M_{\rm P}^4} +\frac{1}{12}\ln\frac{\rho_r}{M_{\rm P}^4}.
\end{eqnarray}
\end{itemize}
It is clear that the second case depends on the reheat temperature due to the existence of the term $\rho_r$ in the expression itself. It is concluded that equation \eqref{efolds3} can be calculated by taking different values of $T_{reh}$ so that one can see the relation between reheat temperature and the number of e-folds  $N_*$, as well as its effects on inflationary predictions. Throughout the next section, we will present the inflationary predictions of the Einstein frame $\beta$-exponential potential with minimal coupling in Palatini $R^2$ gravity supposing the standard thermal history afterward inflation. We first present our results analytically with \textit{rough} approximations as an example, as well as we show the inflationary predictions numerically for both first (instant reheating) and second ($T_{reh}=10^8$ GeV and $T_{reh}=10^{14}$ GeV) cases for the number of e-folds, $N_*$ that we describe above.

\section{Results and Discussion}
\label{results}

\begin{figure}[t!]
	\centering
	\includegraphics[angle=0, width=12.1cm]{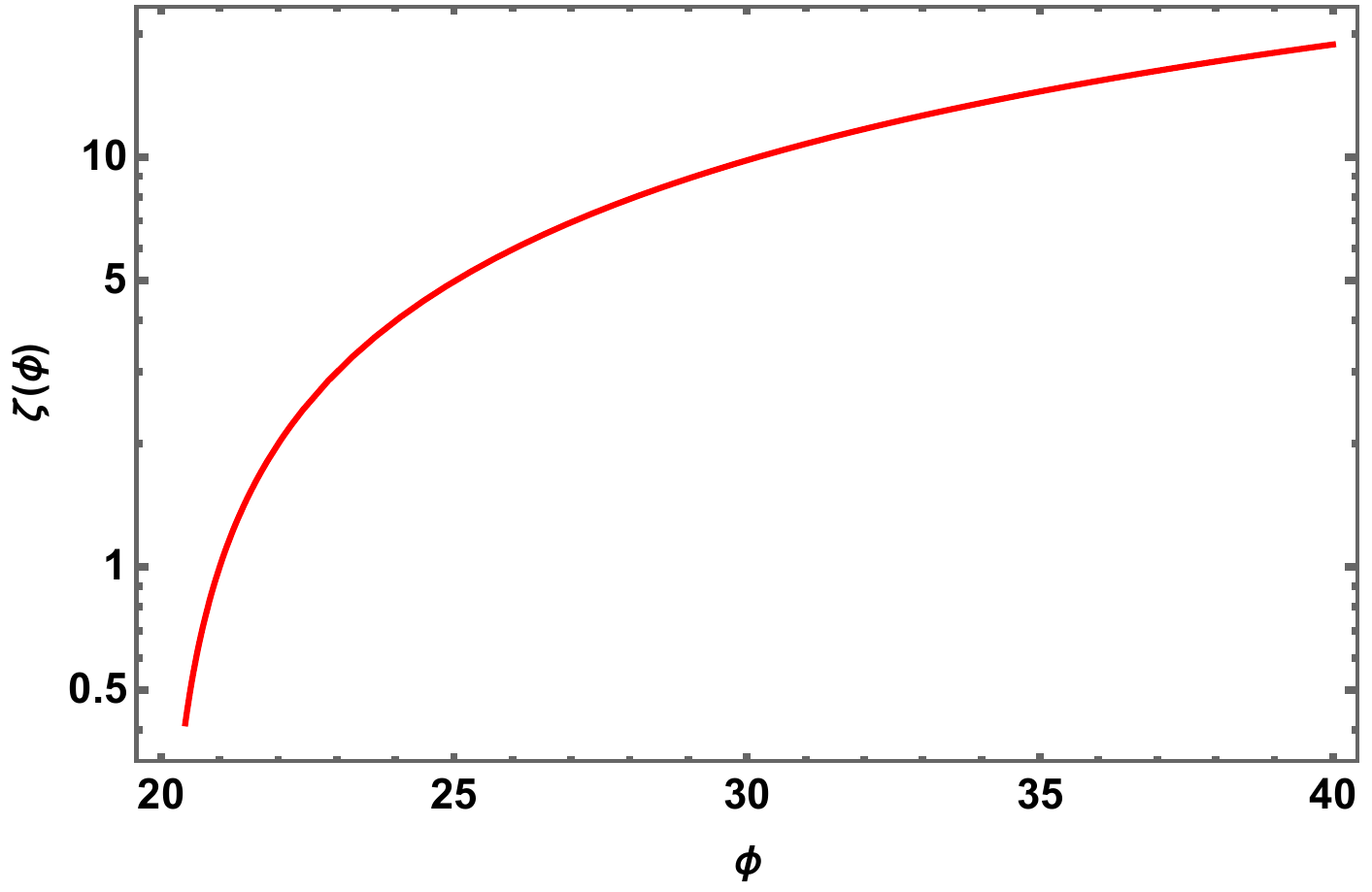}
	\caption{Plot of $\zeta (\phi)$  vs. $\phi$. We set the values as follows: $V_0 = 10^{-9}$, $\alpha=10^8$, $\lambda = 0.1$, $\beta=0.5$ ($M_{\rm P}$ is set to 1). 
}
	\label{fig:zetaphi}
\end{figure}
We inaugurate this section by expressing the canonically normalized field $\zeta$ with respect to the original scalar field $\phi$. By using eq. \eqref{scalardefinit}, the canonical scalar field $\zeta (\phi)$ can be found for the $\beta$-exponential inflation, which is given in eq. \eqref{jframebetamodel} as follows:

\begin{equation}
\mathrm{d}\zeta = \frac{\mathrm{d}\phi}{\sqrt{1 + \frac{4\alpha V_0}{M_{\rm P}^4} \left(1 - \lambda \beta \frac{\phi}{M_{\rm P}}\right)^{1/\beta}}},
\end{equation}

to able to compute and solve this differential equation, we integrate both sides:

\begin{equation}
\zeta(\phi) = \int \frac{\mathrm{d}\phi}{\sqrt{1 + \frac{4\alpha V_0}{M_{\rm P}^4} \left(1 - \lambda \beta \frac{\phi}{M_{\rm P}}\right)^{1/\beta}}}.
\end{equation}

Let \( \gamma \equiv 1 - \lambda \beta \frac{\phi}{M_{\rm P}} \), then the integral becomes:
\begin{equation}
\zeta(\gamma) = -\frac{M_{\rm P}}{\lambda \beta} \int \frac{\mathrm{d}\gamma}{\sqrt{1 + \frac{4\alpha V_0}{M_{\rm P}^4} \gamma^{1/\beta}}}
\end{equation}
This integral can be evaluated in terms of a hypergeometric function. We obtain the final result by inserting $ \gamma \equiv 1 - \lambda \beta \frac{\phi}{M_{\rm P}}$, as follows:
\begin{eqnarray}\label{canonicfieldpot}
  \zeta(\phi) = \frac{(\beta  \lambda  \phi-M_{\rm P}) \times \, _2F_1\left(\frac{1}{2},\beta ;\beta +1;-\frac{4 V_0 \alpha  \left(1-\frac{\phi \beta  \lambda }{M_{\rm P}}\right)^{1/\beta }}{M_{\rm P}^4}\right)}{\beta  \lambda },
\end{eqnarray}
where ${}_2F_1 (a,b;c;z)$ is the hypergeometric function. In order to illustrate the behavior of the function of $\zeta(\phi)$, we compute it numerically as a function of $\phi$. The resultant plot is given in figure~\ref{fig:zetaphi}, by setting $V_0 = 10^{-9}$, $\alpha=10^8$, $\lambda = 0.1$, $\beta=0.5$ and $M_{\rm P}=1$. 

It is important to emphasize that the expression of our potential model in terms of $\phi(\zeta)$ is not straightforward. As in our case, some inflationary potentials are difficult to express in terms of the canonical scalar field $\zeta$ with the exact form due to the complex structures of the potentials, for this case, the analytical approximations for the inflationary potentials are required. For further analytical approximations for our model in this study, please see the appendix section~\ref{sec: appendix}. In addition, because of this situation, for the general values of free parameters in the potential, we compute the inflationary predictions of our model through the numerical techniques with respect to the original scalar field $\phi$ to find the inflationary observables more precisely.

Next, we analyze the inflationary parameters in the slow-roll approximation. Throughout the subsequent analysis in this section, $M_{\rm P}$ will be set to unity. By using eq. \eqref{slowroll2}, the slow-roll parameters, $\epsilon(\phi_*)$ and $\eta(\phi_*)$, can be found for the Einstein frame minimally coupled $\beta$-exponential potential in Palatini $R^2$ gravity, which is defined in eq. \eqref{eframebetamodel} as follows:
\begin{eqnarray}
    \epsilon(\phi_*) \simeq \frac{\lambda ^2}{8 \alpha  V_0 x^{\frac{1}{\beta }+2}+2 x^2}, \nonumber\\ \eta(\phi_*) \simeq \frac{\lambda ^2 \left(-2 \beta +\frac{3}{4 \alpha  V_0 x^{1/\beta }+1}-1\right)}{2 x^2},
\end{eqnarray}
where $1-\beta \lambda \phi_* \equiv x$. Also, by utilizing eq. \eqref{nsralpha1}, the main observational parameters, the spectral index $n_s$ and the tensor-to-scalar ratio $r$, can be acquired analytically for the $\beta$-exponential potential as follows:
\begin{eqnarray}\label{nsrana}
    n_s (\phi_*) \simeq 1-\frac{(2 \beta +1) \lambda ^2}{x^2}, \ \ r (\phi_*) \simeq \frac{8 \lambda ^2}{x^2 \left(4 \alpha  V_0 x^{1/\beta }+1\right)}.
\end{eqnarray}
Also, by using eq. \eqref{efold2}, the amplitude of the curvature perturbation can be found with the form:
\begin{eqnarray}\label{analypert}
\Delta_\mathcal{R}^2 (\phi_*) \simeq \frac{V_0^3 x^{3/\beta }}{12 \pi ^2 \left(4 \alpha  V_0 x^{1/\beta }+1\right){}^4 \left| \frac{x^{\frac{1}{\beta }-1} \lambda  V_0}{\left(4 \alpha  V_0 x^{1/\beta }+1\right){}^2}\right| {}^2}.
\end{eqnarray}
In our numerical analysis, we take into account the slow-roll conditions to calculate $\phi_e$ by using $\epsilon(\phi_e)=1$ in eq. \eqref{slowroll2}, and to compute $\phi_*$ we use the CMB constraint by using eq. \eqref{efold2} with $\Delta_\mathcal{R}^2(\phi_*)\approx 2.1\times10^{-9}$, as well as with these field values, then we set that $50 \lesssim N_* \lesssim 60$ should be satisfied for both the eq. \eqref{perturb2} and \eqref{efolds}. It is also important to note that in the case of instant reheating, $N_*\sim 55-60$. On the other hand, since the reheat temperature is included in the e-fold expression for the $w_r=0$ case, the number of e-folds varies depending on the reheating temperature. We can highlight that as the reheat temperature decreases, the number of e-folds decreases accordingly. For instance, for $T_{reh} = 10^{8}$ GeV, $N_*\sim 50-55$.

\begin{figure}[t!]
	\centering
	\includegraphics[angle=0, width=8.1 cm]{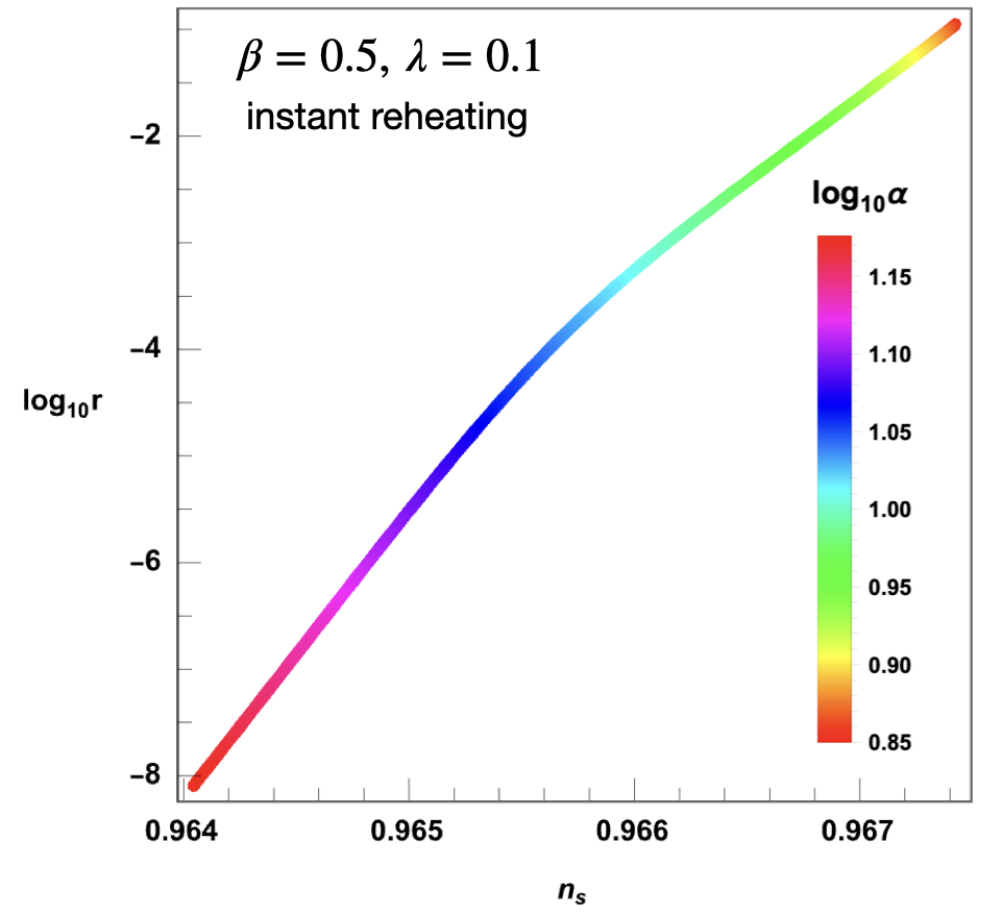}
	\caption{The plot depicts how the $n_s-r$ predictions change depending on the $\alpha$ parameter for the $\beta$-exponential inflation with an $R^2$ term in Palatini formalism with $\beta = 0.5$ and $\lambda = 0.1$ for the instant reheating assumption.
}
	\label{fig:minimalnsralpha}
\end{figure}

Furthermore, regarding eq. \eqref{nsrana}, it is important to mention that for the $\beta$-exponential potential in minimal coupling with an $R^2$ term in Palatini formalism, even though this is not the case for the spectral index $n_s$ predictions, tensor-to-scalar ratio $r$ depends on the $\alpha$ parameter significantly, which is confirmed by our numerical results shown in table~\ref{table:merged}, and related more analysis can be acquired through studying figure~\ref{fig:minimalnsralpha} in a specific range of $\alpha$ values, we will elaborate on this point during the analysis of the related figures. From the analytical results that are given by eq. \eqref{nsrana}, it can be mentioned that the predictions of the tensor-to-scalar ratio $r$ should decrease as the $\alpha$ parameter increases. In addition, the number of e-folds $N_*$ is obtained by using eq. \eqref{perturb2} for our inflationary model in the following form:
\begin{eqnarray}\label{efoldana}
    N_* \simeq \frac{\phi_* (\beta  \lambda  \phi_* -2)}{2 \lambda}.
\end{eqnarray}
One can find the spectral index $n_s$, the tensor-to-scalar ratio $r$, and the amplitude of the curvature perturbation expressions given in eqs. \eqref{nsrana} and \eqref{analypert} in terms of the number of e-folds $N_*$ for the $\beta$-exponential potential by considering different kinds of approximations. For instance, let us assume $\beta \phi_*^2 /2 \gg \phi_* / \lambda$ in eq. \eqref{efoldana}, then we can obtain $x \approx 1 \mp (\lambda \sqrt{2\beta N_*})$. If one inserts this into the equations \eqref{nsrana} and \eqref{analypert}, the predictions can be acquired in terms of the number of e-folds $N_*$ as follows:
\begin{eqnarray}\label{nsrana3}
    n_s \simeq 1-\frac{(2 \beta +1) \lambda ^2}{\left(1 \mp (\lambda \sqrt{2\beta N_*})\right)^2}, \nonumber\\
    r \simeq \frac{8 \lambda ^2}{\left(1 \mp (\lambda \sqrt{2\beta N_*})\right)^2 \left(4 \alpha  V_0 \left(1 \mp (\lambda \sqrt{2\beta N_*})\right)^{1/\beta }+1\right)}, \nonumber\\
\end{eqnarray}
\begin{eqnarray}\label{analypert2}
\Delta_\mathcal{R}^2 \simeq \frac{V_0^3 (1 \mp (\lambda \sqrt{2\beta N_*}))^{3/\beta }}{12 \pi ^2 \left(4 \alpha  V_0 (1 \mp (\lambda \sqrt{2\beta N_*}))^{1/\beta }+1\right){}^4 \left| \frac{(1 \mp (\lambda \sqrt{2\beta N_*}))^{\frac{1}{\beta }-1} \lambda  V_0}{\left(4 \alpha  V_0 (1 \mp (\lambda \sqrt{2\beta N_*}))^{1/\beta }+1\right){}^2}\right| {}^2}.
\end{eqnarray}

\begin{figure*}[t!]
    \centering
    \includegraphics[width=\textwidth]{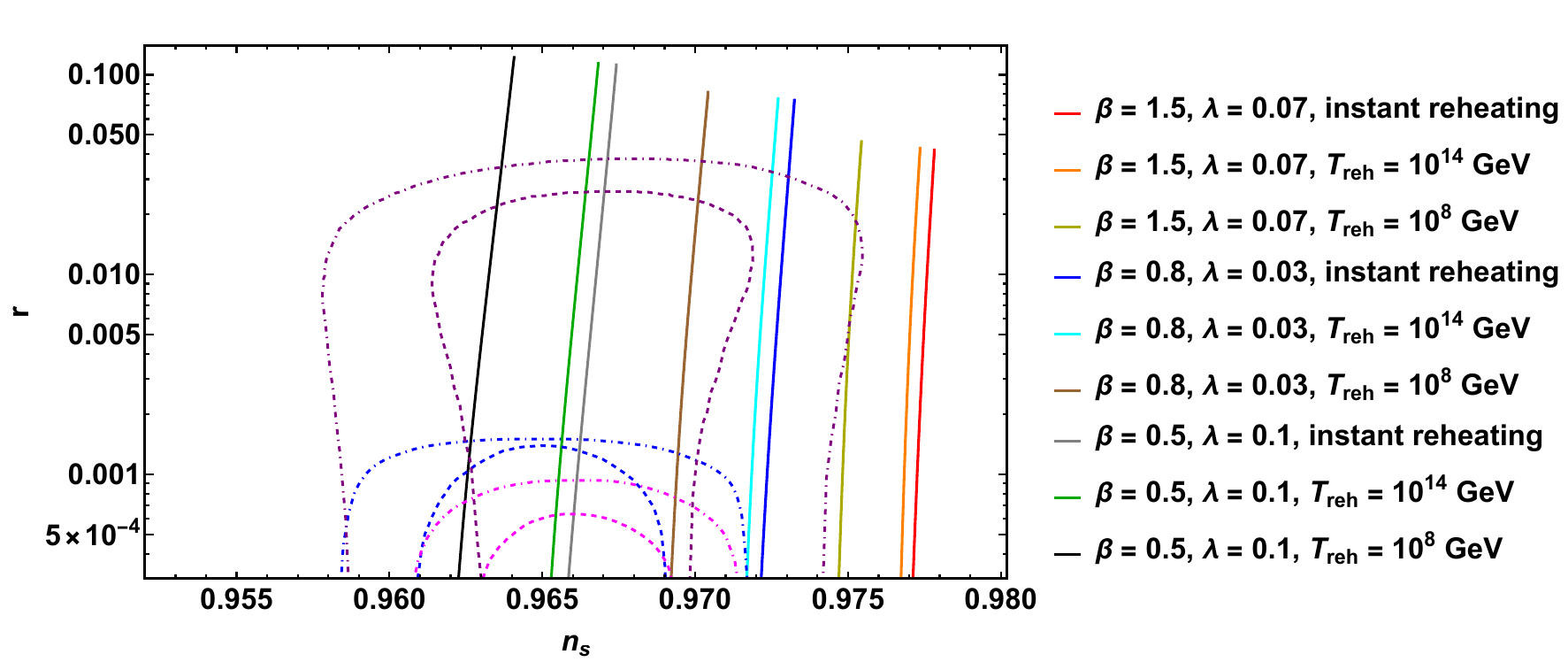}
    \caption{Predictions for the $n_s-r$ parameter space for the selected parameters of minimally coupled $\beta$-exponential inflation with an $R^2$ term in Palatini formalism, where $\alpha$ varies in the range of $[10^7-10^{15}]$. Purple (dot-dashed)(dashed) contours indicate the recent $95\% (68\%)$ CL given by BICEP/Keck \cite{BICEP:2021xfz}, while the magenta (dot-dashed)(dashed) lines correspond to the prospect of future CMB-S4 constraints \cite{Abazajian:2019eic}. Blue (dot-dashed)(dashed) lines represent the $95\% (68\%)$ CL upper limits achievable with LiteBIRD/Planck \cite{LiteBIRD:2022cnt} in the future.}
    \label{fig:minimalnsr}
\end{figure*}

In addition, for the case of $\beta \lambda \phi_* \ll 1 $, one can find $x \approx 1+ \lambda^2 \beta N_* $. With this approximation, the predictions can be obtained in terms of the number of e-folds $N_*$ analytically as follows:
\begin{eqnarray}\label{nsrana4}
    n_s \simeq 1-\frac{(2 \beta +1) \lambda ^2}{(1+ \lambda^2 \beta N_*)^2}, \nonumber\\
    r \simeq \frac{8 \lambda ^2}{(1+ \lambda^2 \beta N_*)^2 \left(4 \alpha  V_0 (1+ \lambda^2 \beta N_*)^{1/\beta }+1\right)}.
\end{eqnarray}

\begin{eqnarray}\label{analypert3}
\Delta_\mathcal{R}^2 \simeq \frac{V_0^3 (1+ \lambda^2 \beta N_*)^{3/\beta }}{12 \pi ^2 \left(4 \alpha  V_0 (1+ \lambda^2 \beta N_*)^{1/\beta }+1\right){}^4 \left| \frac{(1+ \lambda^2 \beta N_*)^{\frac{1}{\beta }-1} \lambda  V_0}{\left(4 \alpha  V_0 (1+ \lambda^2 \beta N_*)^{1/\beta }+1\right){}^2}\right| {}^2}.
\end{eqnarray}

It is also worth emphasizing that these analytical approaches are rough approximations, these are considered for our model in this work for the sake of presenting how the expressions can be written approximately in terms of the number of e-folds in general case. As for the next step, we will begin by discussing our numerical results. It is important to mention that with the assumption of the standard thermal history after inflation, we use eq. \eqref{efolds} for our numerical calculations. Throughout our numerical computations for the inflationary predictions, for each steps, it was also taken into consideration that the values calculated from equation \eqref{efolds} should be very close, almost similar values to the values computed from equation \eqref{perturb2}. The consistency of the results of these two equations ensures that the calculated inflationary parameters are more reliable with high precision.

Our examination in figure~\ref{fig:minimalnsr} takes into consideration different options of the reheating scenarios in order to have a better image of our model's consistency, which leads to a both better and more accurate analysis. We take the highest $T_{reh} = 10^{14}$ GeV, as well as take the lower value, $T_{reh} = 10^{8}$ GeV to show the differences of reheating effects on inflationary predictions, the instant reheating scenario is included as well. Moreover, one can relate the consistency of the plot to the equations derived and presented in this work. As $\epsilon$ decreases, one can see that the predicted values of the tensor-to-scalar ratio $r$ decrease as well, and this leads to lines with different slopes. The positions of the lines in this plot imply the sensitivity of the tensor-to-scalar ratio $r$ and the spectral index $n_s$ to parameters as $\beta, \, \lambda,$ and the reheating scenarios. Comparing the constrained predictions obtained from our model to the observational data, one can spot how our model fits the data right and well, especially given that the solid lines approach the best-fit regions of the contours, which makes our model viable even for the sensitivity forecasts for the future CMB-S4 and LiteBIRD/Planck. Additionally, this figure shows the essential importance of taking into consideration multiple missions, such as CMB-S4 and LiteBIRD/Planck, since both can provide enhanced future sensitivity on the inflationary parameters. As one can observe from this figure, the magenta, and blue data contours represent the $95\%$ and $68\%$ confidence levels for the achievable upper limits in the future; hence, from these missions, we will be able to get further robustness that is in alignment with our model. It is worth mentioning that the consistency with the CMB-S4 and LiteBIRD/Planck predictions highlights the our potential model to be highly viable with the upcoming observational data.

Furthermore, ref.~\cite{dosSantos:2021vis} has examined the $\beta$-exponential inflationary model for both minimally and non-minimally coupled scalar fields with gravity. They have shown a $n_s-r$ plane for both cases. In particular, for the minimally coupled case, they have found their model predictions of $n_s-r$ are in good agreement at $2\sigma$ CL when using Planck 2015 data but for the most recent data from Planck 2018 + BAO measurements, the agreement between their results and the recent data is lost. On the contrary, our results in this work show that the inflationary predictions for the $\beta$-exponential inflation with minimally coupled in Palatini $R^2$ gravity can have a good agreement within the recent Planck + BICEP/Keck data even in $1\sigma$ CL region. Also, for the non-minimal coupling case, ref.~\cite{dosSantos:2021vis} has indicated the agreement between their results and recent cosmological data for some of the selected non-minimal coupling parameters.

\begin{figure}[t!]
	\centering
	\includegraphics[angle=0, width=13.2cm]{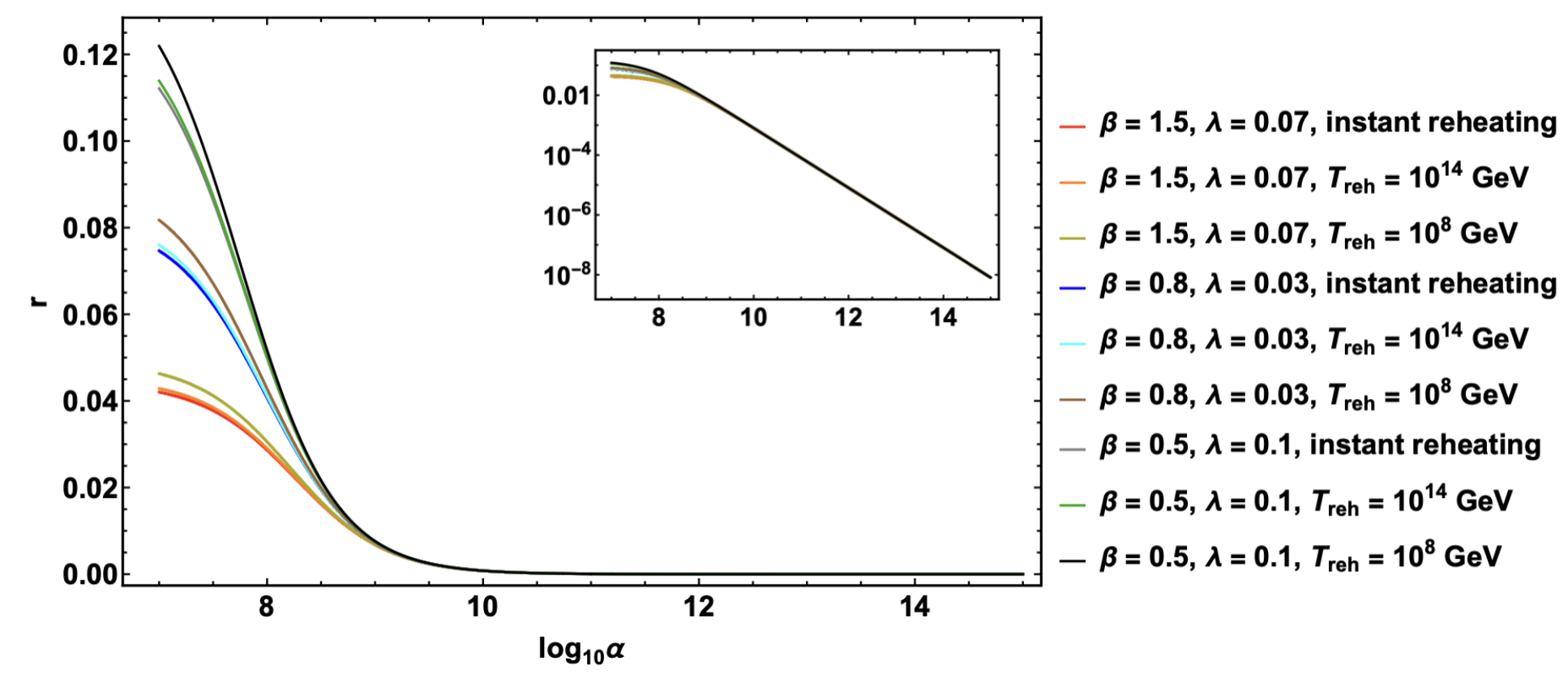}
	\caption{ $\alpha-r$ plane for the selected parameters of minimally coupled $\beta$-exponential inflation with an $R^2$ term in Palatini formalism.
}
	\label{fig:ralphaparam}
\end{figure}

In addition, figure~\ref{fig:ralphaparam} depicts the relation between the tensor-to-scalar ratio $r$ and the parameter $\alpha$ for different values of $\beta,\, \lambda$, and the reheat temperature $T_{reh}$ cases. It can be noticed from this figure that the tensor-to-scalar ratio $r$ decreases as $\alpha$ values increase for all sets of $\beta,\, \lambda,$ and $T_{reh}$. This kind of behavior is consistent with the expectation that for larger $\alpha$ values, we can spot more suppressed tensor modes, resulting in lower tensor-to-scalar ratio $r$ values. On the other hand, for lower values of $\beta$, in a specific choice of $\alpha$, we can notice higher values of the tensor-to-scalar ratio $r$ as well, reflecting in the sensitivity of the inflationary dynamics to this mentioned parameter. Additionally, we cannot ignore the impact of the reheat temperature $T_{reh}$ choice or case as well, for that in some given $\beta$ and $\lambda$ values; instant reheating scenario leads to slightly different tensor-to-scalar ratio $r$ values compared to the scenario of $T_{reh} = 10^8$ GeV. This difference can be interpreted due to the interplay between the reheating phase and the dynamics of the scalar field $\phi$ during inflation.

\begin{figure}[t!]
	\centering
 
    \includegraphics[angle=0, width=12.2cm]{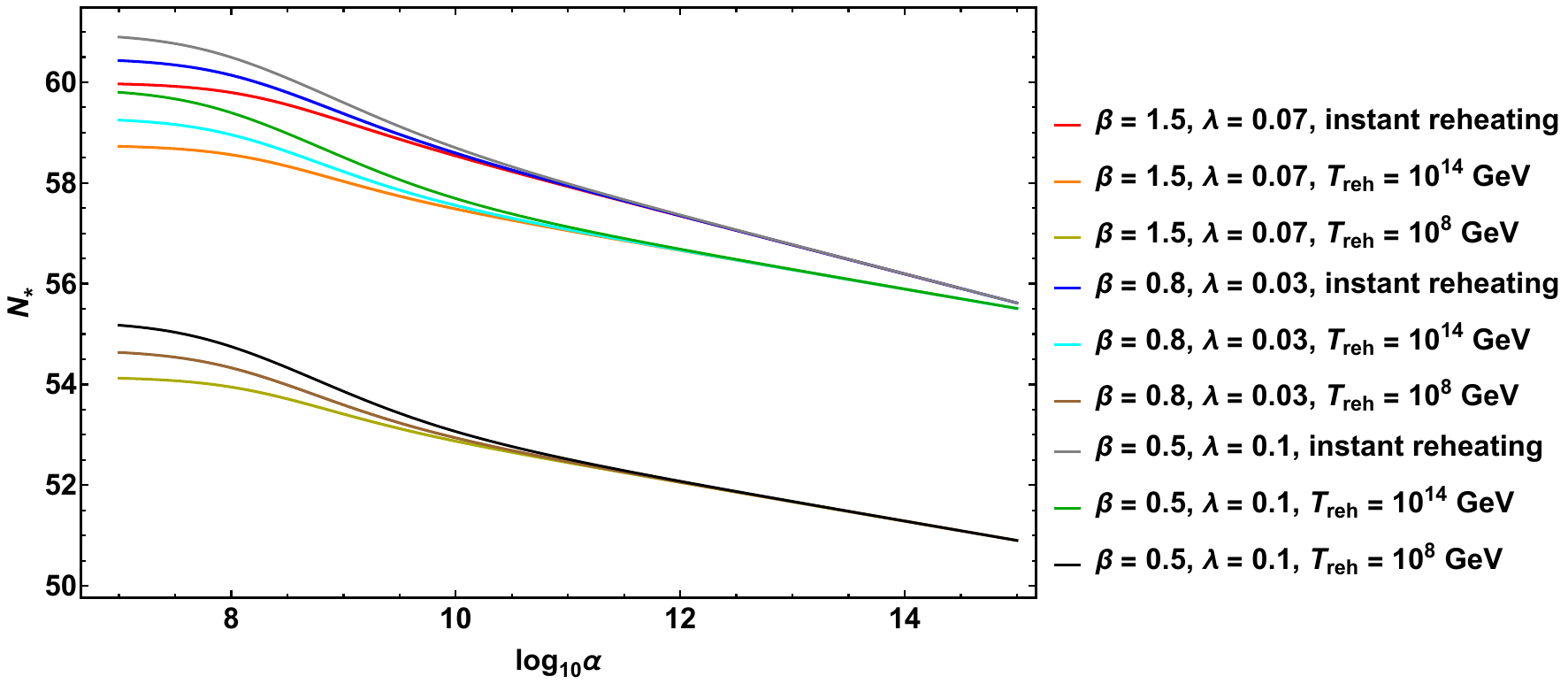}
	\includegraphics[angle=0, width=7.0cm]{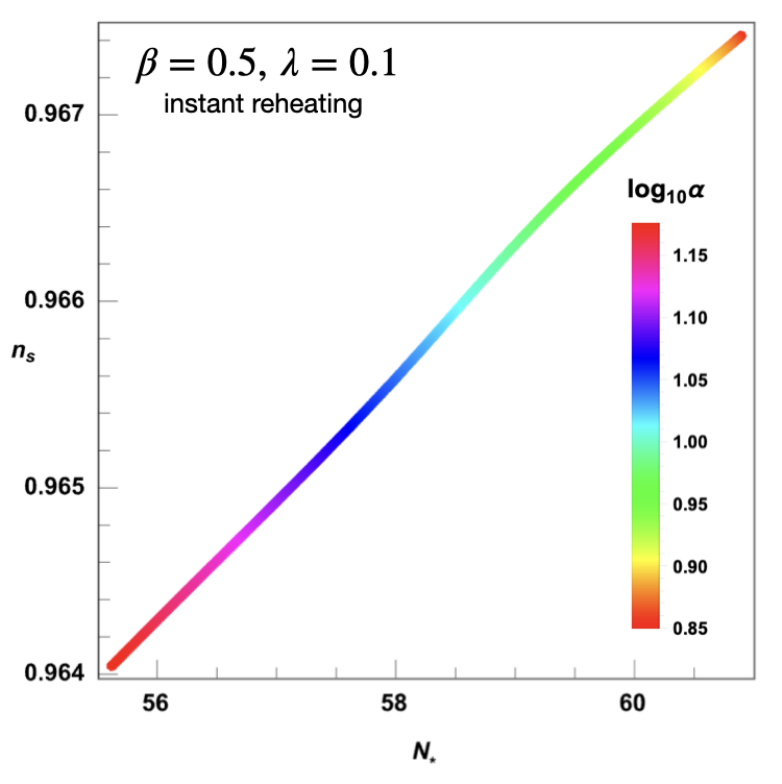}
 \caption{For the minimally coupled $\beta$-exponential inflation with an $R^2$ term in Palatini formalism, the top panel depicts $\alpha-N_*$ plane for the selected parameters, as well as bottom panel shows how the $N_*-n_s$ predictions change depending on the $\alpha$ parameter with $\beta = 0.5$ and $\lambda = 0.1$ for the instant reheating assumption.
}
	\label{fig:nsnnalpha}
\end{figure}

Also, figure~\ref{fig:ralphaparam} shows that for lower values of  $\beta$, the tensor-to-scalar ratio $r$ tends to be higher \textit{an inverse relation}; however, it has a positive correlation with the given values of $\lambda$. This type of behavior of the curves aligns with what one can expect that larger deviations from the standard inflationary potential reflect in a higher tensor-to-scalar ratio $r$. Furthermore, the curves also illustrate the sensitivity of the tensor-to-scalar ratio $r$ to the post-inflationary reheating phase, which is captured by different values of the number of e-folds $N_*$, and this is very important since it points out to the fact that the inflationary predictions depend on the dynamics of the scalar field $\phi$, the slow-roll parameters, and the reheating phase. For the considered scenario of $T_{reh} = 10^{14}$ GeV, the tensor-to-scalar ratio $r$ exhibits slightly higher values when we compare it to the instant reheating case, particularly at lower $\alpha$ values. This behavior further highlights the sensitivity of the inflationary predictions to the reheating phase, where a higher $T_{reh}$ can lead to a more pronounced tensor mode contribution, thus slightly elevating the tensor-to-scalar ratio $r$.

For figure~\ref{fig:nsnnalpha}, we can comment with the following. The top panel presents the relation between the number of e-folds $N_*$ and the parameter $\alpha$ for selected values of $\beta$ and $\lambda$. The figure shows that as the parameter $\alpha$ increases, the number of e-folds $N_*$ decreases for each set of parameters, indicating that a stronger $R^2$ term reduces the duration of inflation. This reduction is more pronounced for lower $\beta$ values, reflecting the dependence of the inflationary dynamics on the shape of the potential, as $\beta$ controls the steepness of the potential. The effect of the dimensionless parameter $\alpha$ in the top plot is subtle and does not play a major role when one compares its relation with the slow-roll parameters, which is stronger as we can notice from the other figures (e.g., figure~\ref{fig:ralphaparam}), and this explains why there is no direct and obvious dependence of the number of e-folds on the parameter $\alpha$ in eq. \eqref{efoldana}. Hence this shows the alignment between our analytical and numerical results, which makes our paper more viable. The bottom panel examines how the spectral index $n_s$ varies with the number of e-folds $N_*$ for the fixed values of $\beta = 0.5$ and $\lambda = 0.1$, under the assumption of instant reheating. The color gradient represents different values of the logarithmic scale of $\alpha$. Additionally, the figure illustrates that as the number of e-folds $N_*$ increases, the spectral index $n_s$ also increases slightly. In the case of $T_{reh} = 10^{14}$ GeV, the spectral index $n_s$ exhibits slightly higher values for the same number of e-folds $N_*$ with comparison to the instant reheating scenario, indicating that a higher reheating temperature can lead to a less tilted curve; hence, affecting the inflationary predictions.


\begin{figure}[t!]
	\centering
	\includegraphics[angle=0, width=13.0cm]{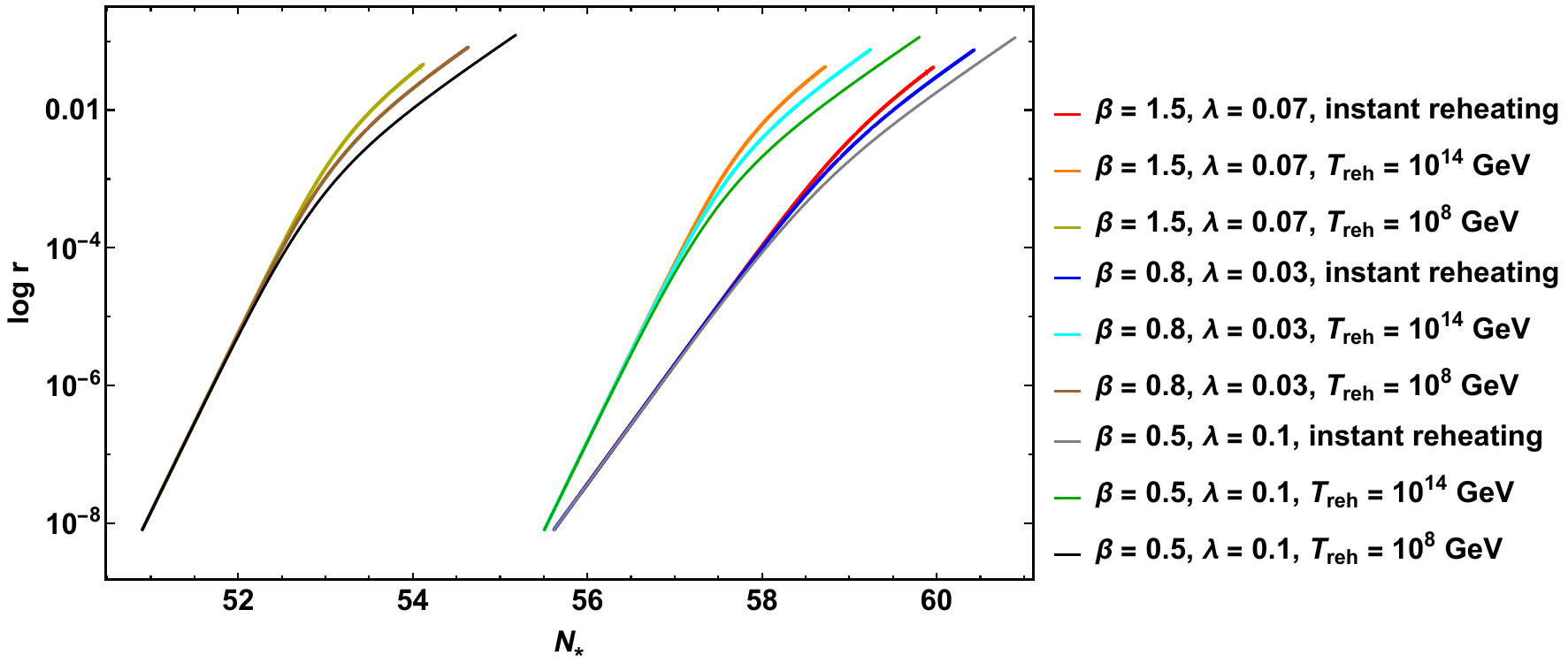}
	\caption{ $N_*-r$ plane for the selected parameters of minimally coupled $\beta$-exponential inflation with an $R^2$ term in Palatini formalism.
}
	\label{fig:rNparam}
\end{figure}

Moreover, figure~\ref{fig:rNparam} represents the tensor-to-scalar ratio $r$ as a function of the number of e-folds $N_*$ for various sets of $\beta$ and $\lambda$. Different curves correspond to the different reheating scenarios. The curves generally show a decreasing trend of the tensor-to-scalar ratio $r$ as the number of e-folds $N_*$ increases. For the higher number of e-folds $N_*$ values, the tensor-to-scalar ratio $r$ tends to stabilize, showing less variation, which is consistent with the expectation that as the inflationary phase progresses, the contributions of the number of e-folds $N_*$ to the tensor-to-scalar ratio $r$ diminish. The comparison between instant reheating and reheating at $ T_{reh} = 10^8$ GeV illustrates that lower reheating temperatures generally lead to slightly lower values of the tensor-to-scalar ratio $r$ for the same number of e-folds $N_*$. This is consistent with the expectation that a longer reheating phase (lower $T_{reh}$) allows for additional red-shifting of the tensor modes, reducing the tensor-to-scalar ratio $r$. As shown from the figure, the scenario at which $T_{reh} = 10^{14}$ GeV results in slightly higher values of the tensor-to-scalar ratio $r$ compared to the lower reheating temperature scenarios, for the same number of e-folds $N_*$. This indicates that a higher reheating temperature reduces the duration of the reheating phase, leading to less red-shifting of tensor modes and thus a higher tensor-to-scalar ratio $r$.

In figure~\ref{fig:minimalnsralpha}, we take advantage of the color gradient technique to indicate the continuous dependence of the tensor-to-scalar ratio $r$ on both the spectral index $n_s$ and the parameter $\alpha$, which provides us with a good alignment with what we have already expected analytically earlier in this section. From this figure, a positive correlation between the tensor-to-scalar ratio $r$ and the spectral index $n_s$ can be noticed, which is something typical of many inflationary models. However, as the spectral index $n_s$ gets larger, the curve starts to behave in a more flat manner, and this can indicate that the sensitivity of the change of the tensor-to-scalar $r$ with respect to the spectral index $n_s$ becomes less sensitive to changes in $\alpha$. One can refer to this last sentence as output to the slow-roll parameters $\epsilon$ and $\eta$ stabilizing in this regime, which is something that can be expected. Looking at the ranges of the tensor-to-scalar ratio $\log_{10} r$, which is produced by our model, we can see their agreement with the observational data.

\renewcommand{\arraystretch}{1.2} 

\begin{table*}[t!]
    \centering
    \resizebox{\textwidth}{!}{%
    \begin{tabular}{c c c c c c c}
        \hline
        $N_*$ & $n_s (\phi_*)$ & $r (\phi_*)$ & $\mathrm{d}n_s/\mathrm{d}\ln k$ ($\phi_*$) & $\phi_{*}$ & $\phi_{e}$ & $V_0$ \\
        \hline
        \multicolumn{7}{c}{$\alpha = 10^8$, $\lambda = 0.1$, $\beta = 0.5$} \\
        \hline
        $61$ & $0.967$ & $0.049$ & $- 5.29 \times 10^{-4}$ & $35.68$ & $21.40$ & $6.58 \times 10^{-9}$ \\
        $55$ & $0.964$ & $0.051$ & $- 6.46 \times 10^{-4}$ & $34.91$ & $21.40$ & $8.04 \times 10^{-9}$ \\
        $45$ & $0.956$ & $0.055$ & $- 9.64 \times 10^{-4}$ & $33.49$ & $21.39$ & $1.20 \times 10^{-8}$ \\
        \hline
        \multicolumn{7}{c}{$\alpha = 10^{15}$, $\lambda = 0.1$, $\beta = 0.5$} \\
        \hline
        $61$ & $0.967$ & $8.04 \times 10^{-9}$ & $- 5.34 \times 10^{-4}$ & $35.64$ & $20.07$ & $6.65 \times 10^{-9}$ \\
        $55$ & $0.964$ & $8.04 \times 10^{-9}$ & $- 6.47 \times 10^{-4}$ & $34.91$ & $20.07$ & $8.05 \times 10^{-9}$ \\
        $45$ & $0.956$ & $8.04 \times 10^{-9}$ & $- 9.64 \times 10^{-4}$ & $33.49$ & $20.06$ & $1.20 \times 10^{-8}$ \\
        \hline
    \end{tabular}
    }
    \caption{The inflationary parameter sets of approximate values for the minimally coupled $\beta$-exponential inflation with an $R^2$ term in Palatini formalism. The table presents essential inflationary observables, the scalar spectral index $n_s$, the tensor-to-scalar ratio $r$, and the running of the spectral index $\mathrm{d}n_s/\mathrm{d}\ln k$ evaluated at the horizon exit of the pivot scale, $\phi_*$. The values of $\phi_*$ and $\phi_e$ correspond to the field values at horizon exit and at the end of inflation, respectively. Two distinct cases for the parameter $\alpha$ are considered: $\alpha = 10^8$ and $\alpha = 10^{15}$, with $\lambda = 0.1$ and $\beta = 0.5$ fixed for both cases. The results for different values of $N_*$, the number of e-folds, are shown to remark the dependence of the inflationary observables (note that the results of predictions in the table were calculated by taking $M_{\rm P}$ set to unity).}
    \label{table:merged}
\end{table*}

Lastly, we also show our results in table~\ref{table:merged}, which is a comprehensive table regarding three different parameters of $\alpha,\, \lambda,$ and $\beta$, and the results presented here have been checked by running the calculations through a numerical technique to obtain a very valid set of results. It is worth mentioning that these choices of the parameter values presented in the table are the ones at which the numerical results are significantly important. From this table, we can mention that for the selected values of e-folds, $45, 55$, and $61$, the spectral index $n_s$ predictions remain similar for the larger and smaller values of $\alpha$ parameter, but the change in the $r$ values is quite large which is something we expect after observing our both results from analytical approximations and figure~\ref{fig:minimalnsralpha}. Thus, it can be concluded that the $\beta$-exponential inflation with an $R^2$ term in Palatini formalism can be well aligned  with the recent cosmological data for the larger $\alpha$ values, which makes the inflationary model compatible with the data. We can see that as the parameter $\alpha$ value increases, the tensor-to-scalar ratio $r$ value becomes very tiny. In addition, from our results, we can mention that the spectral index $n_s$ values generally show a change depending on reheating scenarios. It is pivotal to note that here for the larger number of e-folds $N_*$, larger spectral index $n_s$ values are obtained. Also, the predictions of the running of the spectral index $\mathrm{d}n_s / \mathrm{d}\ln k$ do not alter so much with the $\alpha$ values. Lastly, we find the running of the spectral index $\mathrm{d}n_s / \mathrm{d}\ln k$ predictions are very tiny to be observed at least in the near future.

\section{Summary and conclusions}\label{conc}
In this work, we have studied the minimally coupled $\beta$-exponential inflation with an $R^2$ term in Palatini formalism. For this model, we calculate the inflationary predictions thoroughly, as well as compare them with the recent cosmological data and sensitivity prospects of future CMB experiments: CMB-S4 and LiteBIRD.

The analysis provided in this paper explores the impact of the reheating dynamics, particularly on the spectral index $n_s$, and the tensor-to-scalar ratio $r$, across a variety of reheat temperature scenarios. It is important to remark that higher values of reheat temperature are specifically favorable for non-thermal dark matter production and leptogenesis since higher temperatures increase the chances of interactions and processes that are essential for the occurrence of these phenomena. In addition, we have shown that the inclusion of the reheating impacts gives us clearer information about the effects of the reheating on the inflationary predictions, which not only refines our model's predictions for the inflationary models but also aligns our results closely with the observational constraints from current data, and makes it sensitive for future achievable upper limits forecasts by CMB-S4 and LiteBIRD/Planck, as one can note from figure~\ref{fig:minimalnsr}.

We have found that our considered potential model can be aligned and in good agreement with the recent cosmological data for larger $\alpha$ values, which makes our model consistent with the data, taking very tiny $r$ values, by reaching $r\sim 10^{-9}$. This numerical result is in good agreement with the results we have found analytically for this case. In addition, this result is consistent with the studies in the literature as we discussed in the previous section. We have also discussed our results by considering different reheat temperature scenarios. We show that depending on the values of reheat temperature, the difference appears in the spectral index $n_s$ predictions for the selected parameters in our inflationary model. In addition, we display the predictions of the running of the spectral index $\mathrm{d}n_s / \mathrm{d}\ln k$ are too small for our model.

In addition, it is important to note that the studies in literature for the $\beta$-exponential inflation models that are taking into consideration the different scenarios and gravity models are still ongoing, thus it is considered that our results in this work will be very important to be depicted to build a bridge between different inflationary scenarios, gravity theories and braneworld cosmological frameworks regarding this type of potential model. The $\beta$-exponential potential models can be obtained through the braneworld scenario framework, and considering these kinds of potential models within the context of the braneworld scenarios is strongly motivated for the primordial inflation. Additionally, it is essential to notice that the potential described in this work is an effective one, which becomes truncated when the arguments of the $\beta$-exponential function turn negative. This kind of truncation reflects the limit of the effective description, as the model does not describe a physical relevance at this point. Therefore, this limitation should be highly considered when interpreting the behavior of the potential, specifically when analyzing the implications for the early universe dynamics. 

Last but not least, while the analysis followed in this paper is based around the $\beta$-exponential model, the methodology and the analysis we employed for the predictions of CMB observables and the reheating dynamics are generic and can be applied to a broad range of classes for the inflationary models. For this and the good alignment of our model with the observables, one can agree on the good viability of the model in this paper.

\appendix
\section{Analytical approximations of $\zeta(\phi)$}
\label{sec: appendix}

In this section, we provide a more detailed analysis for finding the inverse relation of the expression $\zeta(\phi)$, which is mentioned with further text in eq. \eqref{canonicfieldpot}. The main challenge comes from the involvement of the hypergeometric function $_2F_1$, which complicates the computation of finding the inversion of $\zeta(\phi)$. Hence, we provide an approximation to find this inversion by employing different approaches to illustrate this issue mathematically, then we depict the results with figures that help us in getting a grasp of our analytical results. 

Moreover, this appendix contains a detailed examination of how the field $\phi$ evolves with respect to $\zeta$, which is essential for understanding the slow-roll parameters and the overall inflationary phase. Through obtaining $\phi(\zeta)$ with the related approximations, we can better analyze the behavior of the potential for the slow-roll regime, which is important for evaluating the number of e-folds $N_*$ as mentioned in details in section \ref{results}, and the circumstances under which the inflation ends.

To find the inverse \(\phi(\zeta)\) analytically from the expression for \(\zeta(\phi)\), we employ equation \eqref{canonicfieldpot}:
\begin{equation}
\zeta(\phi) = \frac{(\beta \lambda \phi - M_{\rm P}) \times \, _2F_1\left(\frac{1}{2}, \beta; \beta + 1; -\frac{4 V_0 \alpha \left(1 - \frac{\phi \beta \lambda}{M_{\rm P}}\right)^{1/\beta}}{M_{\rm P}^4}\right)}{\beta \lambda}. \label{analytzeta}
\end{equation}

It is important to mention here finding the inverse \(\phi(\zeta)\) analytically is challenging because this expression involves the hypergeometric function ${}_2F_1 (a,b;c;z)$, which is not easily invertible in the closed form. However, we can proceed as follows: as the initial step, we express \(\phi\) as a function of \(\zeta\). 

For clarity, let us define:
\begin{equation} \label{an_nu}
\nu \equiv \frac{\beta \lambda \phi}{M_{\rm P}},
\end{equation}

then, the equation \eqref{analytzeta} becomes:
\begin{equation}
\zeta(\nu) = \frac{M_{\rm P}(\nu - 1) \times _2F_1\left(\frac{1}{2}, \beta; \beta + 1; -\frac{4 \alpha V_0 (1 - \nu)^{1/\beta}}{M_{\rm P}^4}\right)}{\beta \lambda}. \label{zeta_nu}
\end{equation}

Since the equation for \(\zeta(\nu)\) is not easily invertible in closed form due to the presence of the hypergeometric function \cite{Maths}, which can be defined by power series for $|z| < 1$ as follows:
\begin{equation}
_2F_1(a,b;c;z) = 1 + \frac{ab}{c} z + \frac{a(a+1)b(b+1)}{2c(c+1)} z^2 + \cdots \, ,
\end{equation}

and using this series expansion, we can approximate:
\begin{equation}\label{the_term}
_2F_1\left(\frac{1}{2}, \beta; \beta + 1; -\frac{4 \alpha V_0 (1 - \nu)^{1/\beta}}{M_{\rm P}^4}\right) \approx 1 - \frac{2 \alpha V_0 \beta (1 - \nu)^{1/\beta}}{M_{\rm P}^4 (\beta + 1)} + \mathcal{O}\left(\frac{(1 - \nu)^{2/\beta}}{M_{\rm P}^8}\right).
\end{equation}

Substituting this approximation back into the expression of \(\zeta(\nu)\) in eq. \eqref{zeta_nu}, we can get the form for the leading-order term as follows
\begin{equation}
\zeta(\nu) \approx \frac{M_{\rm P} (\nu - 1) \left[1 - \frac{2 \alpha V_0 \beta (1 - \nu)^{1/\beta}}{M_{\rm P}^4 (\beta + 1)}\right]}{\beta \lambda}. \label{approxform}
\end{equation}

Now, we aim to invert this equation for \(\nu(\zeta)\), which corresponds to finding \(\phi(\zeta)\). Let us first isolate the leading-order term in \(\nu\) by neglecting higher-order terms in the series expansion:
\begin{equation}
\zeta(\nu) \approx \frac{M_{\rm P} (\nu - 1)}{\beta \lambda}.
\end{equation}

Recall that \(\nu \equiv \frac{\beta \lambda \phi}{M_{\rm P}}\) from eq. \eqref{an_nu}, so we can now solve for \(\zeta(\nu)\) in terms of \(\phi\):

\begin{equation}\label{zeta_phi}
\zeta(\phi) \approx \frac{M_{\rm P} \left(\frac{\beta \lambda \phi}{M_{\rm P}} - 1\right)}{\beta \lambda}.
\end{equation}

Now after we used the previous analytical approximations; starting by eq. \eqref{the_term}, we can easily invert \( \zeta(\phi)\) in eq. \eqref{zeta_phi}, and then we get the following expression:

\begin{equation} \label{phi_zeta_eq}
\phi(\zeta) \approx \frac{M_{\rm P}}{\beta \lambda} + \zeta(\phi).
\end{equation}

To improve this approximation, we can include the next term from the expansion of the hypergeometric function as expressed in eq. \eqref{approxform}.   

\begin{figure}[t!]
	\centering
	\includegraphics[angle=0, width=12.1cm]{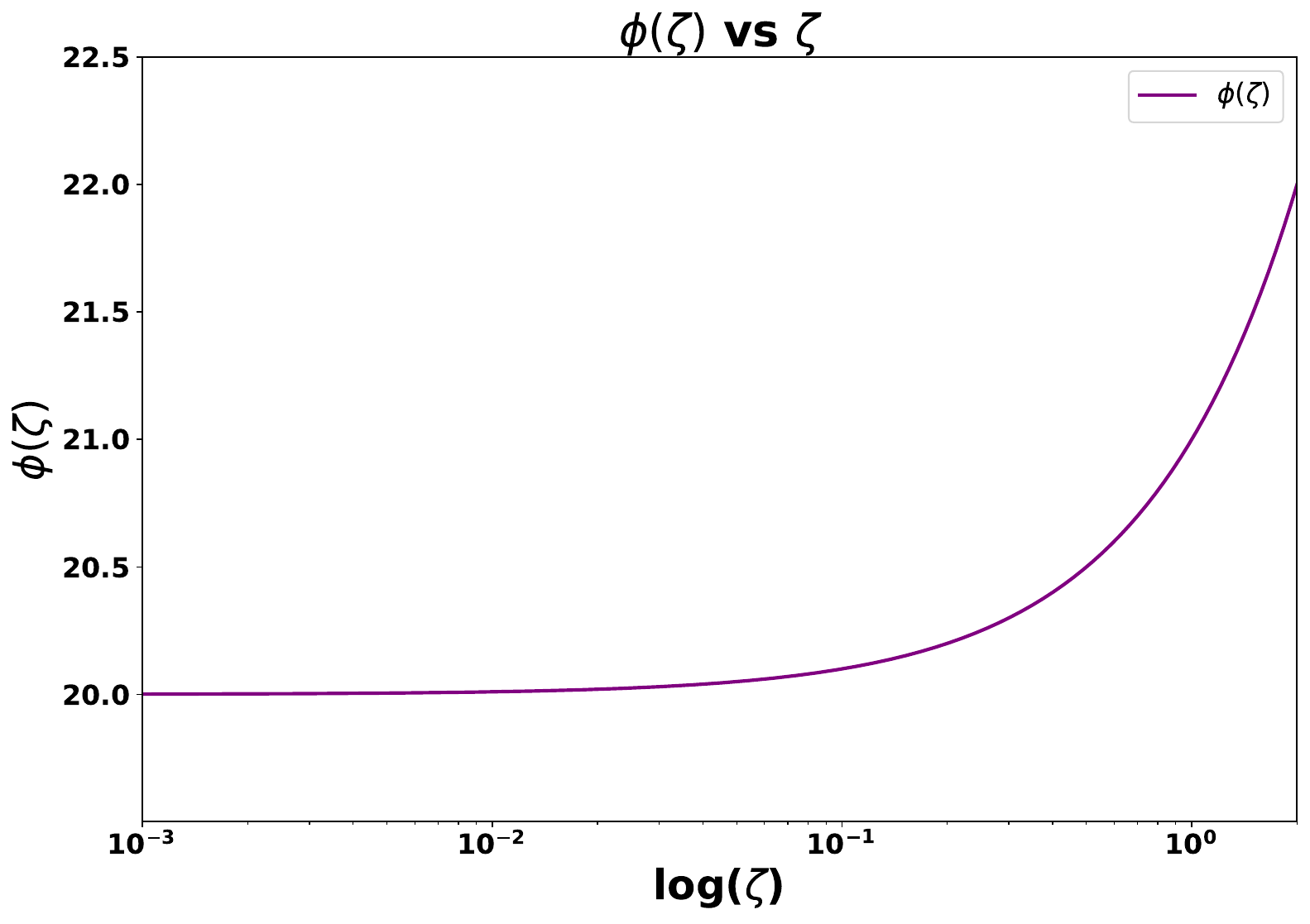}
	\caption{Plot of $\phi (\zeta)$ vs. $\phi$. We set the values as follows: $V_0 = 10^{-9}$, $\alpha=10^8$, $\lambda = 0.1$, $\beta=0.5$ ($M_{\rm P}$ is set to 1).}
	\label{fig:phizeta}
\end{figure}

Expanding \(\nu(\zeta)\) to account for this term requires solving a more complicated equation, which may need numerical inversion. However, the leading-order approximation given above provides a reasonable analytical form for \(\phi(\zeta)\) resulting in the same expression as eq. \eqref{phi_zeta_eq}. Higher-order corrections can be computed by including terms from the hypergeometric series expansion, but these generally require numerical inversion methods. 

Furthermore, we use eq. \eqref{phi_zeta_eq}  to obtain the figure \ref{fig:phizeta} by illustrating the field $\phi(\zeta)$ as a function of the parameter $\zeta$ approximately within the context of $\beta$-exponential potential. The figure points out a characteristic slow-roll inflationary behavior. At lower $\zeta$ values, $\phi(\zeta)$ remains nearly constant, indicating a stable and prolonged inflationary phase, which is essential for generating a sufficient number of e-folds $N_*$, and this is in favor of the analysis presented throughout the paper. This flat region corresponds to the slow-roll approximation, where the potential is sufficiently flat to satisfy inflation. As $\zeta$ increases beyond a certain point, the field begins to grow rapidly, showing the larger field value at the beginning of inflation then the field values will get lower by the slow rolling scenario as mentioned in details and calculations in subsection \ref{inflationary_observables} and section \ref{results}. We highlight that we are just plotting the expression $\phi(\zeta)$ which is computed through rough approximations, so the resultant figure's main purpose is to give us a small hint about the dynamics of the investigated expression of $\phi(\zeta)$.

We can now write the potential in the Einstein frame by substituting eq. \eqref{phi_zeta_eq} into eq. \eqref{eframebetamodel} resulting into getting the potential in terms of $\zeta$ instead of $\phi$, and this can be achieved by following steps:

\begin{equation}
    V_E(\zeta) = \frac{V_0 \left(1 - \lambda \beta \frac{\frac{M_{\rm P}}{\beta \lambda} + \zeta}{M_{\rm P}}\right)^{1/\beta}}{\left(1 + 4 \alpha \frac{V_0 \left(1 - \lambda \beta \frac{\frac{M_{\rm P}}{\beta \lambda} + \zeta}{M_{\rm P}}\right)^{1/\beta}}{M_{\rm P}^4}\right)}. 
\end{equation}

\begin{figure}[t!]
    \centering
    \includegraphics[angle=0, width=11.7cm]{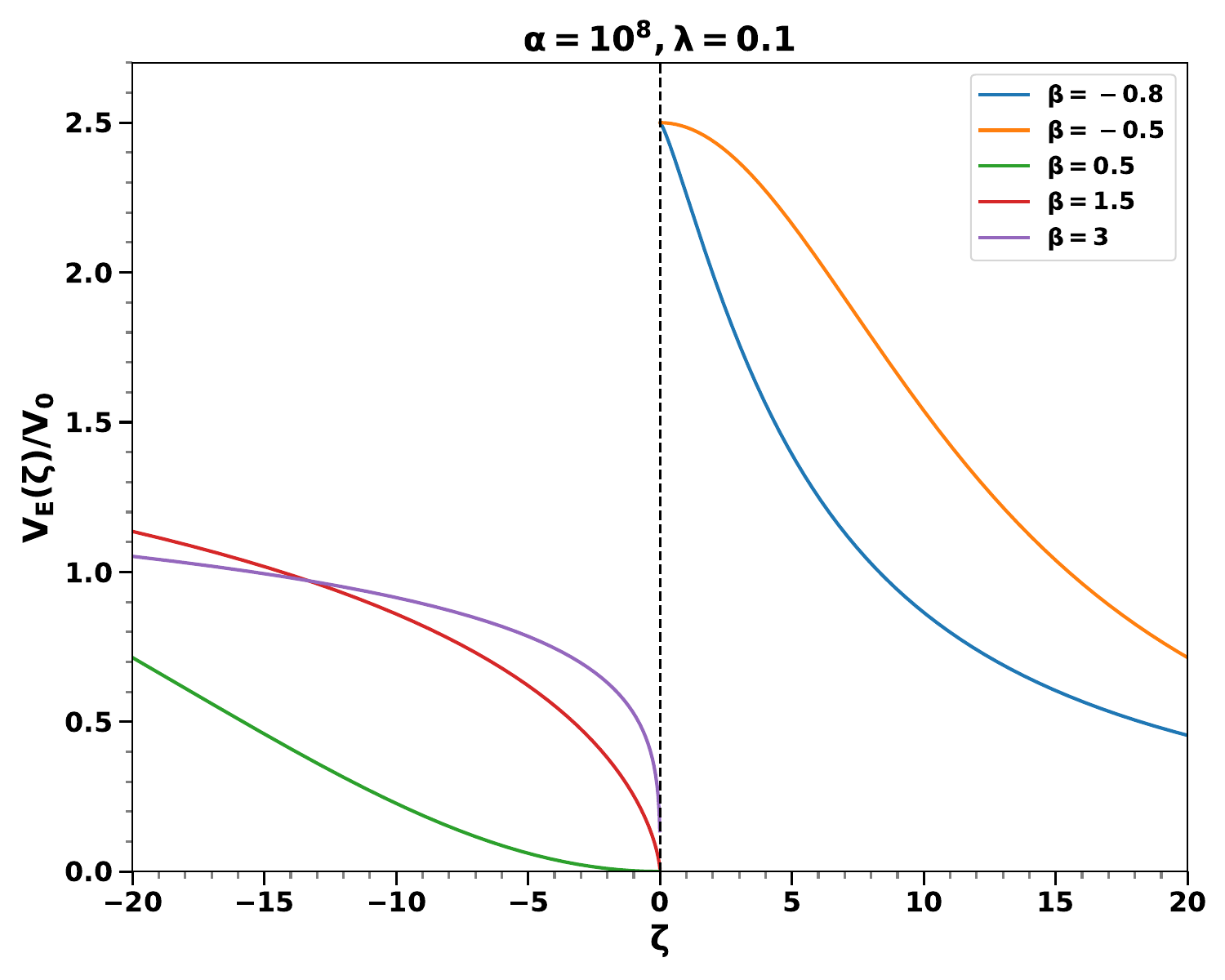}
    \caption{The Einstein frame $\beta$-exponential potential with minimal coupling in Palatini $R^2$ gravity as a function of $\zeta$. The colors show different values of the $\beta$ parameter. We fixed $\lambda = 0.1$, $\alpha = 10^{8}$, $V_0=10^{-9}$, as well as taken $M_{\rm P}$ unity.} 
    \label{fig:Eframepot_zeta}
\end{figure}
\begin{figure}[t!]
    \centering
    \includegraphics[angle=0, width=14.0cm]{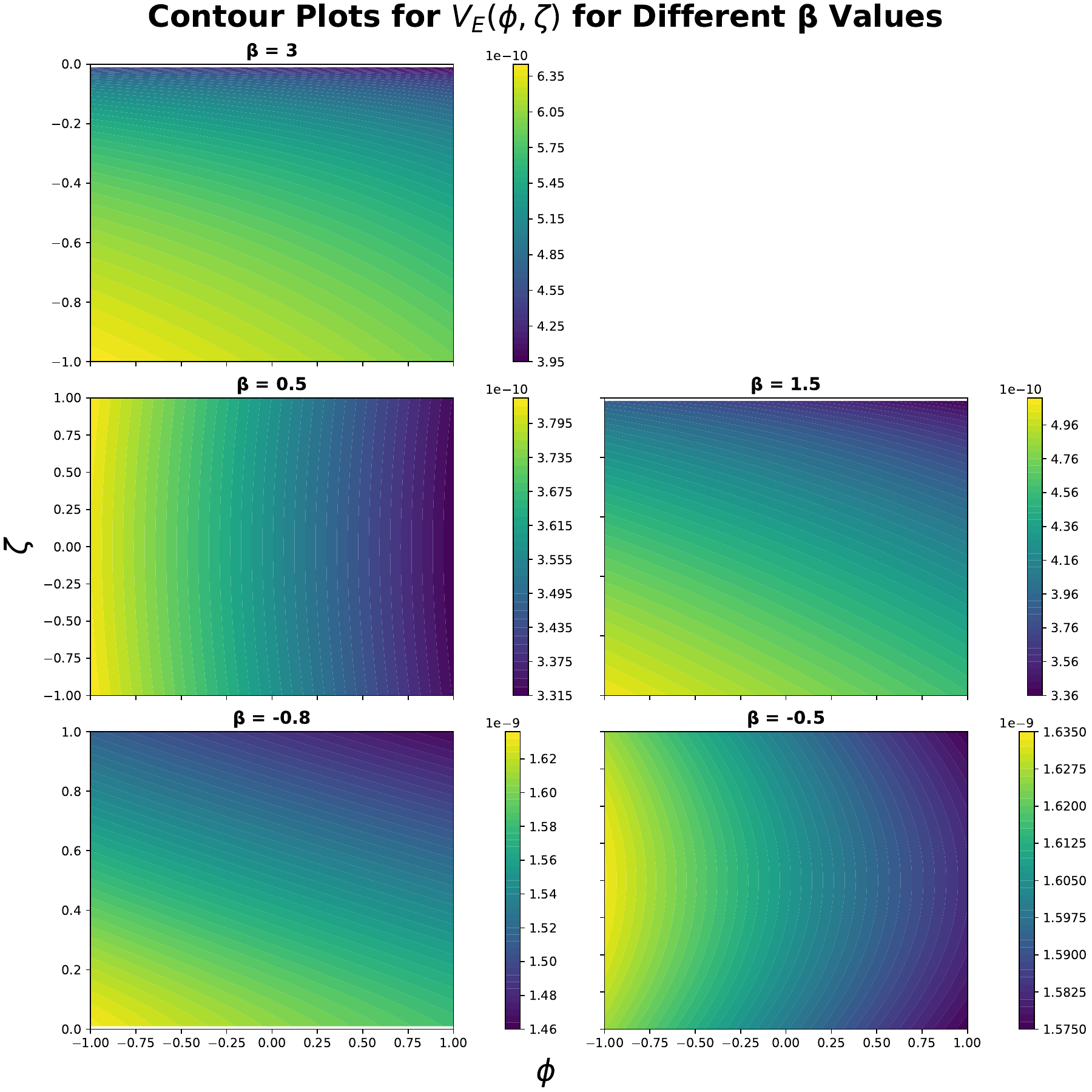}
    \caption{Contour plots of $V_E(\phi, \zeta)$; which are calculated as the average of both our expressed potential in terms of $\phi$ as in eq. (\ref{eframebetamodel}), and in terms of $\zeta$ as shown in eq. (\ref{eframebetamodel_zeta}). The different panels of the plot show the potential behavior for various values of $\beta$: $-0.8,\, -0.5, \, 0.5,\, 1.5,$ and $3$. We fixed $\lambda = 0.1$, $\alpha = 10^{8}$, $V_0=10^{-9}$, as well as taken $M_{\rm P}$ unity.} 
    \label{fig:contour}
\end{figure}
Simplifying the argument of the potential through taking: 
\begin{equation} \label{V_zeta}
    1 - \lambda \beta \frac{\phi}{M_{\rm P}} = 1 - \lambda \beta \left(\frac{\frac{M_{\rm P}}{\beta \lambda} + \zeta}{M_{\rm P}}\right) = - \lambda \beta \frac{\zeta}{M_{\rm P}},
\end{equation}
thus, the expression for the potential $V_E(\zeta)$ can be written as follows
\begin{equation}  \label{eframebetamodel_zeta}
    V_E(\zeta) = \frac{V_0 \left( - \lambda \beta \frac{\zeta}{M_{\rm P}}\right)^{1/\beta}}{\left(1 + 4 \alpha \frac{V_0 \left( - \lambda \beta \frac{\zeta}{M_{\rm P}}\right)^{1/\beta}}{M_{\rm P}^4}\right)},
\end{equation}
which is depicted in figure \ref{fig:Eframepot_zeta}\footnote{It is worth mentioning that the curves we are considering in this figure are the ones with the physical meaning and consistent with our model definition starting by eq. \eqref{potexpgdef}. } in order to illustrate the behavior of the potential in the Einstein frame in terms of $\zeta$ for a variety of choices of $\beta$. It is worth keeping in mind that this plot represents only an approximation of computing $V_E(\zeta)$ analytically and it does not represent the full behavior of our potential, so this plot is solely an attempt to visualize and give a general image of our potential in terms of $\zeta$.

We try to visualize that regardless of the different representations of our potential, the main behavior of the potential itself is retained. We express our potential in terms of $\phi$ as in eq. \eqref{eframebetamodel}. Additionally, in this section, after following an analytical approach and approximations, we are able to represent our potential itself in terms of $\zeta$ as can be seen in eq. \eqref{eframebetamodel_zeta}, so having these two representations we consider analyzing their behavior in both the frequency and the field domains closely. Along this paper, we expressed our potential $V_E$ in terms of $\phi$ and $\zeta$, and to explore the consistency and the behavior of our model within these different representations we illustrate in figure \ref{fig:contour} the contour $V_E(\phi, \zeta)$; which is defined by taking the average of the sum of our potentials' different representations, and this step helps us to illustrate how both $V_E(\phi)$ and $V_E(\zeta)$ are behaving similarly. From this figure, one can notice that regardless of the transformation between $\phi$ and $\zeta$ our potential maintains its expected structure and dynamics ensuring that our model behaves consistently under these different descriptions. This is a solid step to confirm the robustness of the inflationary dynamics through the potential in this paper, which is essential for ensuring accurate cosmological predictions.

\begin{figure}[t!]
	\centering
 
    \includegraphics[angle=0, width=10.0cm]{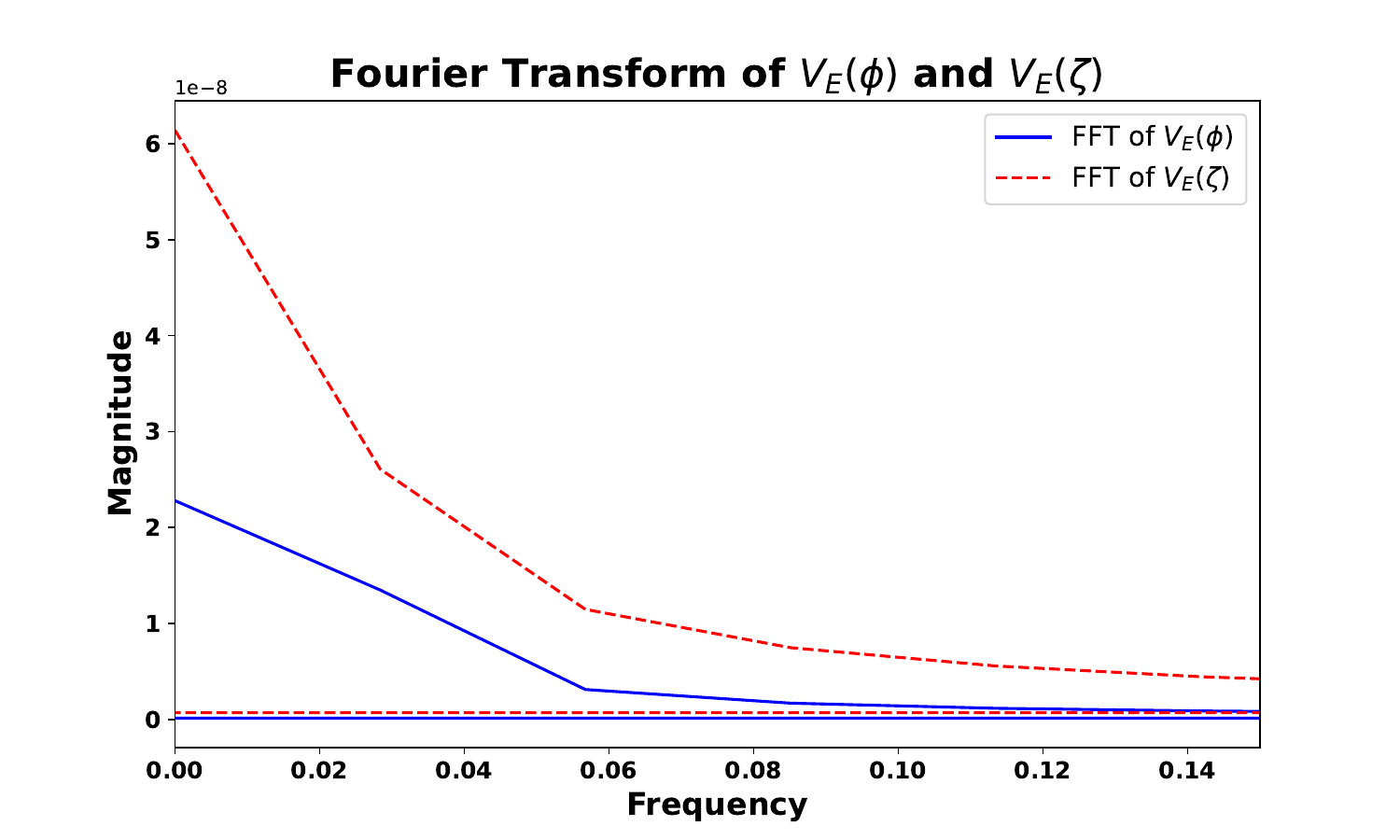}
	\includegraphics[angle=0, width=10.0cm]{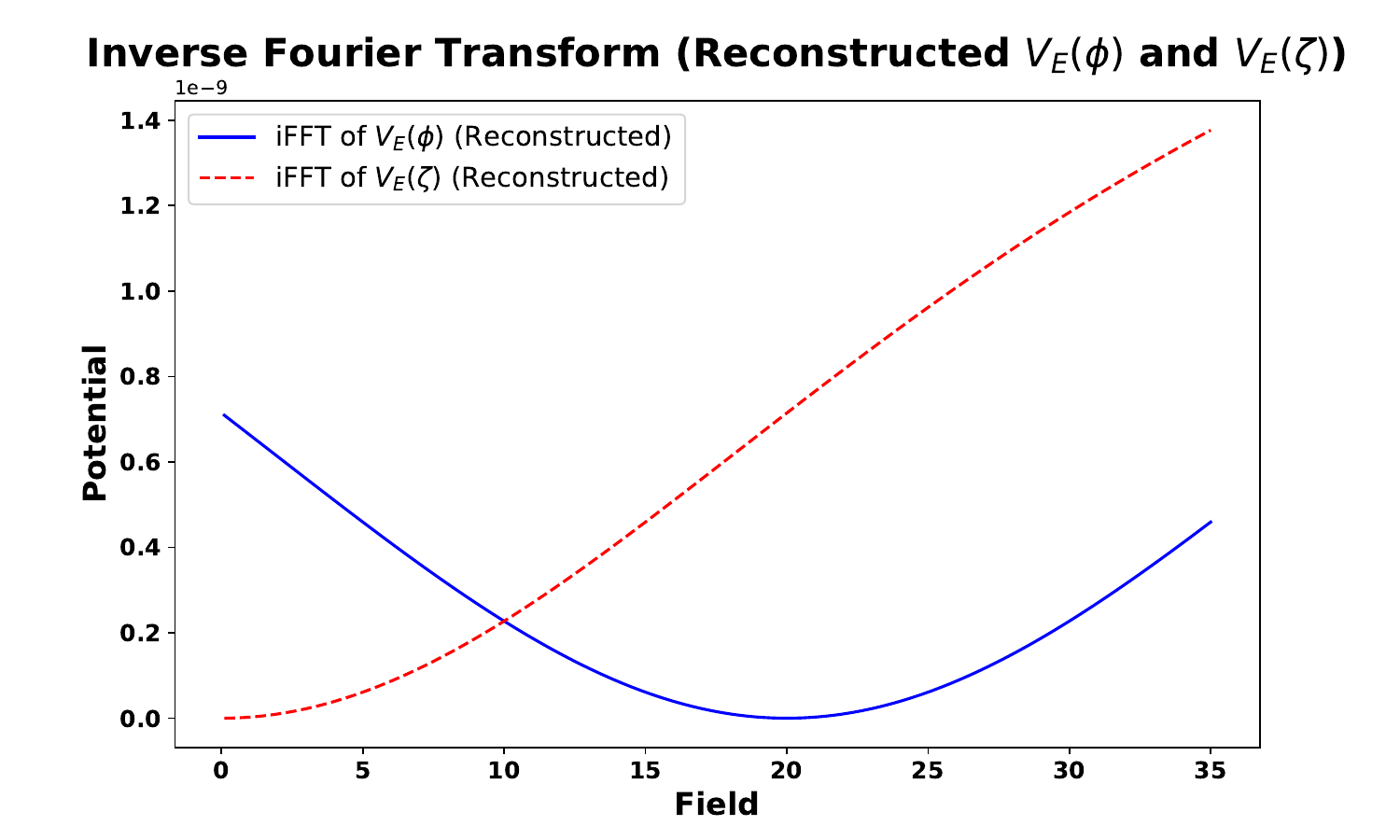}
 \caption{The \textit{top} panel represents the Fourier Transform (FFT) of the potentials $V_E(\phi)$  (solid blue line) and $V_E(\zeta)$ (dashed red line), highlighting their magnitude spectra in the low-frequency domain. Additionally, both the horizontal lines represent the corresponding average value of each potential. The plot at the \textit{bottom} panel shows the reconstructed potentials, with the same color definitions of the top plot, obtained via the inverse Fourier Transform (iFFT). We fixed $\beta = 0.5$, $\lambda = 0.1$, $\alpha = 10^{8}$, $V_0=10^{-9}$, as well as taken $M_{\rm P}$ unity.} 
	\label{fig:transform}
\end{figure}

Moreover, we harness the Fourier transform which gives us the chance to break down these potentials into their frequency components, providing insights into their structure and the impact of different frequency modes. In the same plot, we can observe the two linear lines that represent a very smooth varying near the $x$ axis, and they represent the low-frequency elements that correspond to the overall trend of the two different representations of our potential resulting in no sharp oscillations or rapid changes. From the \textit{top} panel of figure \ref{fig:transform}\footnote{Since this plot is solely representing an additional viability check, it is sufficient to take $\beta$ = 1/2.}, one can notice that the overall frequency content of both potentials is similar; however, due to the approximations that are used to get the expression of $\phi(\zeta)$ we can spot subtle differences, particularly at lower frequencies. For the \textit{bottom} panel of the same figure, we plot the inverse Fourier transform, which depicts the original form of the potentials. Regardless of the approximations used in this section, both $V_E(\phi)$ and $V_E(\zeta)$ follow the same general trends; both of them grow over a range of the field but with different profiles. This shows us that, despite the used approximations we retain the key similarities to $V_E(\phi)$, which indicates that the approximations that are used in calculating $\phi(\zeta)$ affect finer details without drastically altering the overall structure of our potential.

We can also evaluate $\zeta(\phi)$ in eq. \eqref{canonicfieldpot} for the case of $\beta = 1/2$ by utilizing the special case for the hypergeometric function ${}_2F_1 \left( \frac{1}{2}, \frac{1}{2}; \frac{3}{2}; -\Lambda^2 \right)$\footnote{When dealing with functions as $\sinh^{-1}$, one needs to be extra cautious about its relevant physical representation. In order to make sure that our related calculations to this formula are well considered according to this latter property, we need to consider that our term $\Lambda$ is such that $|\Lambda| \leq 1$ which is indeed here in our case represented as $|\Lambda| \ll 1$. Another viability check is by mapping: $\Lambda \rightarrow - i \, \tau$ for any $\tau$, then $|\tau| \leq 1$, which satisfied for our $\Lambda$ choice in this appendix as well \cite{abramowitz1968handbook}. }:
\begin{equation}
    {}_2F_1 \left( \frac{1}{2}, \frac{1}{2}; \frac{3}{2}; -\Lambda^2 \right) = \frac{\sinh^{-1}(\Lambda)}{\Lambda}, 
\end{equation}
using this identity, where $\Lambda = \sqrt{\frac{4 \alpha V_0 \left(1 - \frac{\phi \lambda}{2 M_{\rm P}}\right)^2}{M_{\rm P}^4}}$, the expression for $\zeta(\phi)$ defined by eq. \eqref{canonicfieldpot} becomes:

\begin{equation}
    \zeta(\phi) = \frac{\left( \frac{\lambda \phi - 2 M_{\rm P}}{2} \right) M_{\rm P}^2 \sinh^{-1} \left( \frac{2 \sqrt{\alpha V_0} \left(1 - \frac{\phi \lambda}{2 M_{\rm P}}\right)}{M_{\rm P}^2}\right)}{\lambda \sqrt{\alpha V_0} \left( 1 - \frac{\phi \lambda}{2 M_{\rm P}} \right)}.
\end{equation}

To find \(\phi(\zeta)\), we start by isolating the term involving \(\phi\). Multiply both sides by \(\lambda \sqrt{\alpha V_0} \left( 1 - \frac{\phi \lambda}{2 M_{\rm P}} \right)\):

\begin{equation}
    \zeta(\phi) \cdot \lambda \sqrt{\alpha V_0} \left( 1 - \frac{\phi \lambda}{2 M_{\rm P}} \right) = \frac{(\lambda \phi - 2 M_{\rm P})}{2} M_{\rm P}^2 \sinh^{-1} \left( \frac{2 \sqrt{\alpha V_0} \left(1 - \frac{\phi \lambda}{2 M_{\rm P}}\right)}{M_{\rm P}^2}\right),
\end{equation}

then dividing by \(\frac{\lambda \phi - 2 M_{\rm P}}{2}\), it gives:

\begin{equation}
    \frac{\zeta(\phi) \cdot \lambda \sqrt{\alpha V_0} \left( 1 - \frac{\phi \lambda}{2 M_{\rm P}} \right)}{\frac{\lambda \phi - 2 M_{\rm P}}{2}} = M_{\rm P}^2 \sinh^{-1} \left( \frac{2 \sqrt{\alpha V_0} \left(1 - \frac{\phi \lambda}{2 M_{\rm P}}\right)}{M_{\rm P}^2}\right),
\end{equation}

dividing both sides by \(M_{\rm P}^2\), the expression results in: 

\begin{equation}
    \frac{2 \zeta(\phi) \cdot \lambda \sqrt{\alpha V_0} \left( 1 - \frac{\phi \lambda}{2 M_{\rm P}} \right)}{(\lambda \phi - 2 M_{\rm P}) M_{\rm P}^2} = \sinh^{-1} \left( \frac{2 \sqrt{\alpha V_0} \left(1 - \frac{\phi \lambda}{2 M_{\rm P}}\right)}{M_{\rm P}^2}\right).
\end{equation}
\begin{figure}[t!]
	\centering
	\includegraphics[angle=0, width=12.1cm]{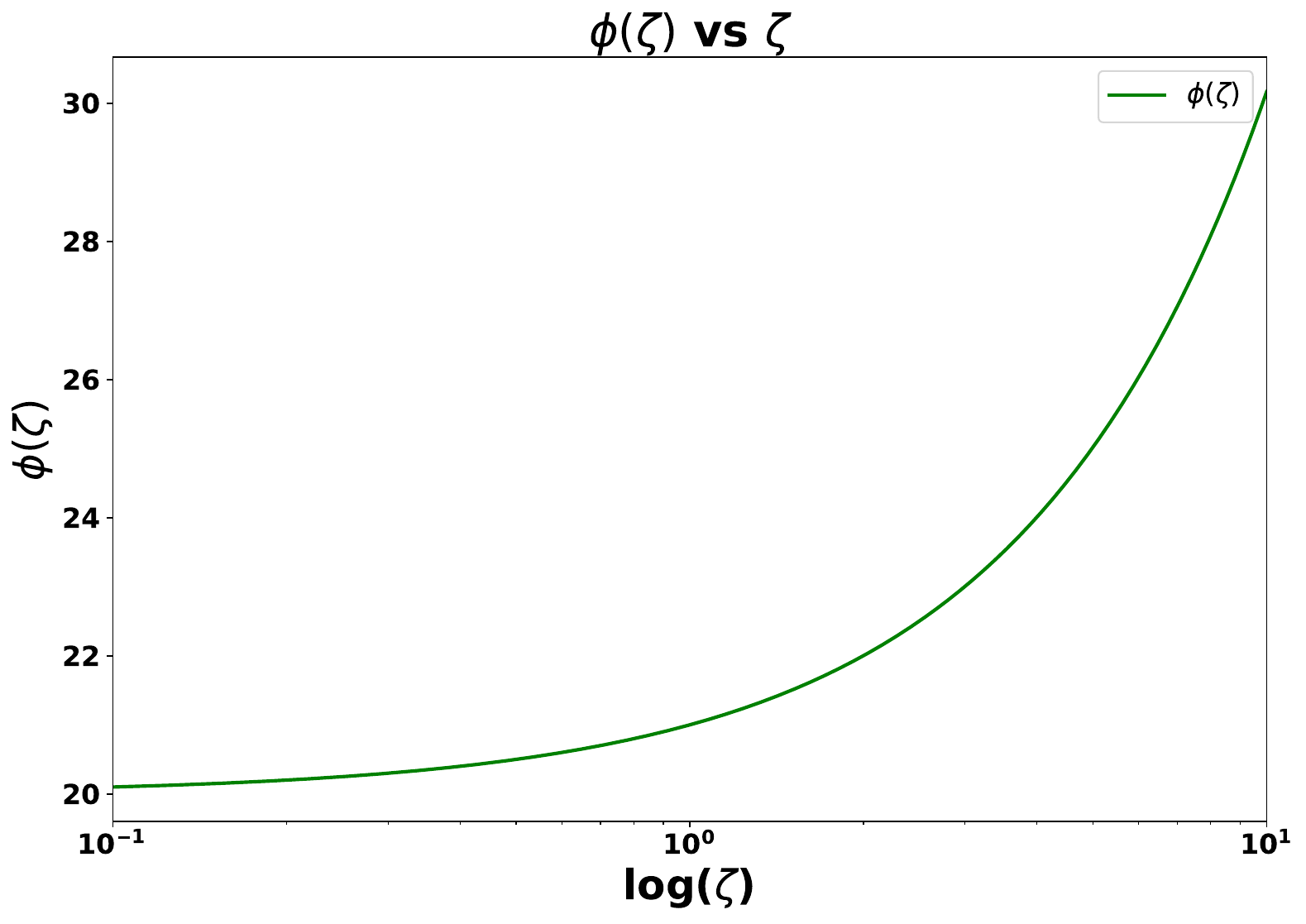}
	\caption{Plot of $\phi (\zeta)$ vs. $\phi$ through using the hypergeometric hyperbolic sine formula for the special case of $\beta = \frac{1}{2}$. We set the values as follows: $V_0 = 10^{-9}$, $\alpha=10^8$, $\lambda = 0.1$ ($M_{\rm P}$ is set to 1).}
	\label{fig:phizeta_numerical}
\end{figure}

By applying the hyperbolic sine function (\(\sinh\)) to both sides, we get:

\begin{equation}
    \frac{2 \sqrt{\alpha V_0} \left(1 - \frac{\phi \lambda}{2 M_{\rm P}}\right)}{M_{\rm P}^2} = \sinh \left( \frac{2 \zeta(\phi) \cdot \lambda \sqrt{\alpha V_0} \left( 1 - \frac{\phi \lambda}{2 M_{\rm P}} \right)}{(\lambda \phi - 2 M_{\rm P}) M_{\rm P}^2} \right),
\end{equation}

then solving this result for \(1 - \frac{\phi \lambda}{2 M_{\rm P}}\), it gives us the following result

\begin{equation}
    1 - \frac{\phi \lambda}{2 M_{\rm P}} = \frac{M_{\rm P}^2}{2 \sqrt{\alpha V_0}} \sinh \left( \frac{2 \zeta(\phi) \cdot \lambda \sqrt{\alpha V_0} \left( 1 - \frac{\phi \lambda}{2 M_{\rm P}} \right)}{(\lambda \phi - 2 M_{\rm P}) M_{\rm P}^2} \right).
\end{equation}

Finally, solving for \(\phi(\zeta)\) provides us with the following expression:

\begin{equation}\label{phizeta_2}
    \phi(\zeta) = \frac{2 M_{\rm P}}{\lambda} \left( 1 - \frac{M_{\rm P}^2}{2 \sqrt{\alpha V_0}} \sinh \left( \frac{2 \zeta(\phi) \cdot \lambda \sqrt{\alpha V_0} \left( 1 - \frac{\phi \lambda}{2 M_{\rm P}} \right)}{(\lambda \phi - 2 M_{\rm P}) M_{\rm P}^2} \right) \right).
\end{equation}

Thus, the inverse $\phi(\zeta)$ involves solving a transcendental equation, which in general cannot be expressed in a simple closed form, which highlights the intricacy of the underlying dynamics in this inflationary model; one of the points that support to this latter point is that we used eq. \eqref{phi_zeta_eq} in plotting $\phi(\zeta)$ vs. $\zeta$ rather than using eq. \eqref{phizeta_2}.  The equation for $\phi(\zeta)$ must be solved numerically or through iterative methods, and this behavior is essential for our understanding of the period of inflation and the conditions under which inflation ends, making the numerical solution very essential for analyzing our scenario for $\beta$-exponential inflation. In conclusion, through different approaches and approximations we formulated two expressions of the invert $\phi(\zeta)$; however, it is worth mentioning that both resultant expressions describe the very same dynamics within our model framework. The expression provided by eq. \eqref{phizeta_2} is more intricacy; however, for the sake of showing the viability and consistency of this expression with the first one in eq. \eqref{phi_zeta_eq}, and plotted in figure \ref{fig:phizeta}, we solve eq. \eqref{phizeta_2} numerically through the Root-finding method \cite{Press2007} and then plot it as shown in figure \ref{fig:phizeta_numerical}. Consequently, both of the figures for the two expressions show the consistency of the behavior of the two given expressions of $\phi(\zeta)$.

\section*{Acknowledgements}
The authors would like to thank Alberto Salvio, Antonio Racioppi and Anish Ghoshal for the useful comments and suggestions to improve the manuscript. NB also thanks Ozan Sarg{\i}n for the fruitful suggestions on the hypergeometric function.

\paragraph{Note added.} Both authors have contributed \textit{equally} to this work.



\begin{thebibliography}{10}

\bibitem{Linde:2007fr}
A.D.~Linde, \emph{{Inflationary Cosmology}}, \href{https://doi.org/10.1007/978-3-540-74353-8\_1}{\emph{Lect. Notes Phys.} {\bfseries 738} (2008) 1} [\href{https://arxiv.org/abs/0705.0164}{{\ttfamily 0705.0164}}].

\bibitem{Starobinsky:1979ty}
A.A.~Starobinsky, \emph{{Spectrum of relict gravitational radiation and the early state of the universe}}, {\emph{JETP Lett.} {\bfseries 30} (1979) 682}.

\bibitem{Linde:1983gd}
A.D.~Linde, \emph{{Chaotic Inflation}}, \href{https://doi.org/10.1016/0370-2693(83)90837-7}{\emph{Phys. Lett. B} {\bfseries 129} (1983) 177}.

\bibitem{Mukhanov:1981xt}
V.F.~Mukhanov and G.V.~Chibisov, \emph{{Quantum Fluctuations and a Nonsingular Universe}}, {\emph{JETP Lett.} {\bfseries 33} (1981) 532}.

\bibitem{Guth:1980zm}
A.H.~Guth, \emph{{The Inflationary Universe: A Possible Solution to the Horizon and Flatness Problems}}, \href{https://doi.org/10.1103/PhysRevD.23.347}{\emph{Phys. Rev. D} {\bfseries 23} (1981) 347}.

\bibitem{Starobinsky:1980te}
A.A.~Starobinsky, \emph{{A New Type of Isotropic Cosmological Models Without Singularity}}, \href{https://doi.org/10.1016/0370-2693(80)90670-X}{\emph{Phys. Lett. B} {\bfseries 91} (1980) 99}.

\bibitem{Martin:2013tda}
J.~Martin, C.~Ringeval and V.~Vennin, \emph{{Encyclop\ae{}dia Inflationaris}}, \href{https://doi.org/10.1016/j.dark.2014.01.003}{\emph{Phys. Dark Univ.} {\bfseries 5-6} (2014) 75} [\href{https://arxiv.org/abs/1303.3787}{{\ttfamily 1303.3787}}].

\bibitem{Benetti:2016jhf}
M.~Benetti and R.O.~Ramos, \emph{{Warm inflation dissipative effects: predictions and constraints from the Planck data}}, \href{https://doi.org/10.1103/PhysRevD.95.023517}{\emph{Phys. Rev. D} {\bfseries 95} (2017) 023517} [\href{https://arxiv.org/abs/1610.08758}{{\ttfamily 1610.08758}}].

\bibitem{SantosdaCosta:2017ctv}
S.~Santos~da Costa, M.~Benetti and J.~Alcaniz, \emph{{A Bayesian analysis of inflationary primordial spectrum models using Planck data}}, \href{https://doi.org/10.1088/1475-7516/2018/03/004}{\emph{JCAP} {\bfseries 03} (2018) 004} [\href{https://arxiv.org/abs/1710.01613}{{\ttfamily 1710.01613}}].

\bibitem{Benetti:2019kgw}
M.~Benetti, L.~Graef and R.O.~Ramos, \emph{{Observational Constraints on Warm Inflation in Loop Quantum Cosmology}}, \href{https://doi.org/10.1088/1475-7516/2019/10/066}{\emph{JCAP} {\bfseries 10} (2019) 066} [\href{https://arxiv.org/abs/1907.03633}{{\ttfamily 1907.03633}}].

\bibitem{SantosdaCosta:2020dyl}
S.~Santos~da Costa, M.~Benetti, R.M.P.~Neves, F.A.~Brito, R.~Silva and J.S.~Alcaniz, \emph{{Brane inflation and the robustness of the Starobinsky inflationary model}}, \href{https://doi.org/10.1140/epjp/s13360-020-01015-1}{\emph{Eur. Phys. J. Plus} {\bfseries 136} (2021) 84} [\href{https://arxiv.org/abs/2007.09211}{{\ttfamily 2007.09211}}].

\bibitem{Misner1973}
C.W.~{Misner}, K.S.~{Thorne} and J.A.~{Wheeler}, \emph{{Gravitation}} (1973).

\bibitem{Wald:1984rg}
R.M.~Wald, \emph{{General Relativity}}, Chicago Univ. Pr., Chicago, USA (1984), \href{https://doi.org/10.7208/chicago/9780226870373.001.0001}{10.7208/chicago/9780226870373.001.0001}.

\bibitem{Ferraris:1982wci}
M.~Ferraris, M.~Francaviglia and C.~Reina, \emph{{Variational formulation of general relativity from 1915 to 1925 \textquotedblleft{}Palatini's method\textquotedblright{} discovered by Einstein in 1925}}, \href{https://doi.org/10.1007/BF00756060}{\emph{Gen. Rel. Grav.} {\bfseries 14} (1982) 243}.

\bibitem{Will:2018bme}
C.M.~Will, \emph{{Theory and Experiment in Gravitational Physics}}, Cambridge University Press (9, 2018).

\bibitem{Bauer:2008zj}
F.~Bauer and D.A.~Demir, \emph{{Inflation with Non-Minimal Coupling: Metric versus Palatini Formulations}}, \href{https://doi.org/10.1016/j.physletb.2008.06.014}{\emph{Phys. Lett. B} {\bfseries 665} (2008) 222} [\href{https://arxiv.org/abs/0803.2664}{{\ttfamily 0803.2664}}].

\bibitem{Racioppi:2019jsp}
A.~Racioppi, \emph{{Non-Minimal (Self-)Running Inflation: Metric vs. Palatini Formulation}}, \href{https://doi.org/10.1007/JHEP01(2021)011}{\emph{JHEP} {\bfseries 21} (2020) 011} [\href{https://arxiv.org/abs/1912.10038}{{\ttfamily 1912.10038}}].

\bibitem{Bostan:2020pnb}
N.~Bostan, \emph{{Palatini double-well and Coleman-Weinberg potentials with non-minimal coupling}}, \href{https://doi.org/10.1088/1475-7516/2021/04/042}{\emph{JCAP} {\bfseries 04} (2021) 042} [\href{https://arxiv.org/abs/2009.04406}{{\ttfamily 2009.04406}}].

\bibitem{Jarv:2017azx}
L.~J\"arv, A.~Racioppi and T.~Tenkanen, \emph{{Palatini side of inflationary attractors}}, \href{https://doi.org/10.1103/PhysRevD.97.083513}{\emph{Phys. Rev. D} {\bfseries 97} (2018) 083513} [\href{https://arxiv.org/abs/1712.08471}{{\ttfamily 1712.08471}}].

\bibitem{Bostan:2019uvv}
N.~Bostan, \emph{{Non-minimally coupled quartic inflation with Coleman-Weinberg one-loop corrections in the Palatini formulation}}, \href{https://doi.org/10.1016/j.physletb.2020.135954}{\emph{Phys. Lett. B} {\bfseries 811} (2020) 135954} [\href{https://arxiv.org/abs/1907.13235}{{\ttfamily 1907.13235}}].

\bibitem{Bostan:2022swq}
N.~Bostan, \emph{{Non-minimally coupled Natural Inflation: Palatini and Metric formalism with the recent BICEP/Keck}}, \href{https://doi.org/10.1088/1475-7516/2023/02/063}{\emph{JCAP} {\bfseries 02} (2023) 063} [\href{https://arxiv.org/abs/2209.02434}{{\ttfamily 2209.02434}}].

\bibitem{Bostan:2023ped}
N.~Bostan and S.~Roy~Choudhury, \emph{{First constraints on non-minimally coupled Natural and Coleman-Weinberg inflation and massive neutrino self-interactions with Planck+BICEP/Keck}}, \href{https://doi.org/10.1088/1475-7516/2024/07/032}{\emph{JCAP} {\bfseries 07} (2024) 032} [\href{https://arxiv.org/abs/2310.01491}{{\ttfamily 2310.01491}}].

\bibitem{Dioguardi:2022oqu}
C.~Dioguardi, A.~Racioppi and E.~Tomberg, \emph{{Beyond (and back to) Palatini quadratic gravity and inflation}}, \href{https://doi.org/10.1088/1475-7516/2024/03/041}{\emph{JCAP} {\bfseries 03} (2024) 041} [\href{https://arxiv.org/abs/2212.11869}{{\ttfamily 2212.11869}}].

\bibitem{Ghoshal:2024ycp}
A.~Ghoshal, Z.~Lalak, S.~Pal and S.~Porey, \emph{{Post-inflationary leptogenesis and dark matter production: metric versus Palatini formalism}}, \href{https://doi.org/10.1007/JHEP06(2024)038}{\emph{JHEP} {\bfseries 06} (2024) 038} [\href{https://arxiv.org/abs/2401.17262}{{\ttfamily 2401.17262}}].

\bibitem{Koivisto:2005yc}
T.~Koivisto and H.~Kurki-Suonio, \emph{{Cosmological perturbations in the palatini formulation of modified gravity}}, \href{https://doi.org/10.1088/0264-9381/23/7/009}{\emph{Class. Quant. Grav.} {\bfseries 23} (2006) 2355} [\href{https://arxiv.org/abs/astro-ph/0509422}{{\ttfamily astro-ph/0509422}}].

\bibitem{Borunda:2008kf}
M.~Borunda, B.~Janssen and M.~Bastero-Gil, \emph{{Palatini versus metric formulation in higher curvature gravity}}, \href{https://doi.org/10.1088/1475-7516/2008/11/008}{\emph{JCAP} {\bfseries 11} (2008) 008} [\href{https://arxiv.org/abs/0804.4440}{{\ttfamily 0804.4440}}].

\bibitem{Gialamas:2023flv}
I.D.~Gialamas, A.~Karam, T.D.~Pappas and E.~Tomberg, \emph{{Implications of Palatini gravity for inflation and beyond}}, \href{https://doi.org/10.1142/S0219887823300076}{\emph{Int. J. Geom. Meth. Mod. Phys.} {\bfseries 20} (2023) 2330007} [\href{https://arxiv.org/abs/2303.14148}{{\ttfamily 2303.14148}}].

\bibitem{Alcaniz:2006nu}
J.S.~Alcaniz and F.C.~Carvalho, \emph{{Beta-exponential inflation}}, \href{https://doi.org/10.1209/0295-5075/79/39001}{\emph{EPL} {\bfseries 79} (2007) 39001} [\href{https://arxiv.org/abs/astro-ph/0612279}{{\ttfamily astro-ph/0612279}}].

\bibitem{Santos:2017alg}
M.A.~Santos, M.~Benetti, J.~Alcaniz, F.A.~Brito and R.~Silva, \emph{{CMB constraints on $\beta$-exponential inflationary models}}, \href{https://doi.org/10.1088/1475-7516/2018/03/023}{\emph{JCAP} {\bfseries 03} (2018) 023} [\href{https://arxiv.org/abs/1710.09808}{{\ttfamily 1710.09808}}].

\bibitem{dosSantos:2021vis}
F.B.M.~dos Santos, S.~Santos~da Costa, R.~Silva, M.~Benetti and J.~Alcaniz, \emph{{Constraining non-minimally coupled \ensuremath{\beta}-exponential inflation with CMB data}}, \href{https://doi.org/10.1088/1475-7516/2022/06/001}{\emph{JCAP} {\bfseries 06} (2022) 001} [\href{https://arxiv.org/abs/2110.14758}{{\ttfamily 2110.14758}}].

\bibitem{Santos:2022exm}
F.B.M.d.~Santos, R.~Silva, S.S.~da~Costa, M.~Benetti and J.S.~Alcaniz, \emph{{Warm $\beta $-exponential inflation and the swampland conjectures}}, \href{https://doi.org/10.1140/epjc/s10052-023-11329-w}{\emph{Eur. Phys. J. C} {\bfseries 83} (2023) 178} [\href{https://arxiv.org/abs/2209.06153}{{\ttfamily 2209.06153}}].

\bibitem{Capistrano:2024kuc}
A.a.J.S.~Capistrano, R.C.~Nunes and L.A.~Cabral, \emph{{Lower tensor-to-scalar ratio as possible signature of modified gravity}}, \href{https://doi.org/10.1103/PhysRevD.109.123517}{\emph{Phys. Rev. D} {\bfseries 109} (2024) 123517} [\href{https://arxiv.org/abs/2403.13860}{{\ttfamily 2403.13860}}].

\bibitem{Dvali:1998pa}
G.R.~Dvali and S.H.H.~Tye, \emph{{Brane inflation}}, \href{https://doi.org/10.1016/S0370-2693(99)00132-X}{\emph{Phys. Lett. B} {\bfseries 450} (1999) 72} [\href{https://arxiv.org/abs/hep-ph/9812483}{{\ttfamily hep-ph/9812483}}].

\bibitem{Mersini-Houghton:2000pof}
L.~Mersini-Houghton, \emph{{Radion potential and brane dynamics}}, \href{https://doi.org/10.1142/S0217732301004911}{\emph{Mod. Phys. Lett. A} {\bfseries 16} (2001) 1583} [\href{https://arxiv.org/abs/hep-ph/0001017}{{\ttfamily hep-ph/0001017}}].

\bibitem{Randall:1999ee}
L.~Randall and R.~Sundrum, \emph{{A Large mass hierarchy from a small extra dimension}}, \href{https://doi.org/10.1103/PhysRevLett.83.3370}{\emph{Phys. Rev. Lett.} {\bfseries 83} (1999) 3370} [\href{https://arxiv.org/abs/hep-ph/9905221}{{\ttfamily hep-ph/9905221}}].

\bibitem{Goldberger:1999uk}
W.D.~Goldberger and M.B.~Wise, \emph{{Modulus stabilization with bulk fields}}, \href{https://doi.org/10.1103/PhysRevLett.83.4922}{\emph{Phys. Rev. Lett.} {\bfseries 83} (1999) 4922} [\href{https://arxiv.org/abs/hep-ph/9907447}{{\ttfamily hep-ph/9907447}}].

\bibitem{Abbott:1982hn}
L.F.~Abbott, E.~Farhi and M.B.~Wise, \emph{{Particle Production in the New Inflationary Cosmology}}, \href{https://doi.org/10.1016/0370-2693(82)90867-X}{\emph{Phys. Lett. B} {\bfseries 117} (1982) 29}.

\bibitem{Albrecht:1982mp}
A.~Albrecht, P.J.~Steinhardt, M.S.~Turner and F.~Wilczek, \emph{{Reheating an Inflationary Universe}}, \href{https://doi.org/10.1103/PhysRevLett.48.1437}{\emph{Phys. Rev. Lett.} {\bfseries 48} (1982) 1437}.

\bibitem{Kofman:1997yn}
L.~Kofman, A.D.~Linde and A.A.~Starobinsky, \emph{{Towards the theory of reheating after inflation}}, \href{https://doi.org/10.1103/PhysRevD.56.3258}{\emph{Phys. Rev. D} {\bfseries 56} (1997) 3258} [\href{https://arxiv.org/abs/hep-ph/9704452}{{\ttfamily hep-ph/9704452}}].

\bibitem{Chung:1998rq}
D.J.H.~Chung, E.W.~Kolb and A.~Riotto, \emph{{Production of massive particles during reheating}}, \href{https://doi.org/10.1103/PhysRevD.60.063504}{\emph{Phys. Rev. D} {\bfseries 60} (1999) 063504} [\href{https://arxiv.org/abs/hep-ph/9809453}{{\ttfamily hep-ph/9809453}}].

\bibitem{Bassett:2005xm}
B.A.~Bassett, S.~Tsujikawa and D.~Wands, \emph{{Inflation dynamics and reheating}}, \href{https://doi.org/10.1103/RevModPhys.78.537}{\emph{Rev. Mod. Phys.} {\bfseries 78} (2006) 537} [\href{https://arxiv.org/abs/astro-ph/0507632}{{\ttfamily astro-ph/0507632}}].

\bibitem{Dai:2014jja}
L.~Dai, M.~Kamionkowski and J.~Wang, \emph{{Reheating constraints to inflationary models}}, \href{https://doi.org/10.1103/PhysRevLett.113.041302}{\emph{Phys. Rev. Lett.} {\bfseries 113} (2014) 041302} [\href{https://arxiv.org/abs/1404.6704}{{\ttfamily 1404.6704}}].

\bibitem{Munoz:2014eqa}
J.B.~Munoz and M.~Kamionkowski, \emph{{Equation-of-State Parameter for Reheating}}, \href{https://doi.org/10.1103/PhysRevD.91.043521}{\emph{Phys. Rev. D} {\bfseries 91} (2015) 043521} [\href{https://arxiv.org/abs/1412.0656}{{\ttfamily 1412.0656}}].

\bibitem{Cook:2015vqa}
J.L.~Cook, E.~Dimastrogiovanni, D.A.~Easson and L.M.~Krauss, \emph{{Reheating predictions in single field inflation}}, \href{https://doi.org/10.1088/1475-7516/2015/04/047}{\emph{JCAP} {\bfseries 04} (2015) 047} [\href{https://arxiv.org/abs/1502.04673}{{\ttfamily 1502.04673}}].

\bibitem{Hanin:2023ypf}
A.~Hanin, K.~El~Bourakadi, M.~Ferricha-Alami, Z.~Sakhi and M.~Bennai, \emph{{Reheating Mechanism from Tree Level Potential in Standard Cosmology}}, \href{https://doi.org/10.1007/s10773-023-05364-2}{\emph{Int. J. Theor. Phys.} {\bfseries 62} (2023) 143}.

\bibitem{Dolgov:1989us}
A.D.~Dolgov and D.P.~Kirilova, \emph{{ON PARTICLE CREATION BY A TIME DEPENDENT SCALAR FIELD}}, {\emph{Sov. J. Nucl. Phys.} {\bfseries 51} (1990) 172}.

\bibitem{Traschen:1990sw}
J.H.~Traschen and R.H.~Brandenberger, \emph{{Particle Production During Out-of-equilibrium Phase Transitions}}, \href{https://doi.org/10.1103/PhysRevD.42.2491}{\emph{Phys. Rev. D} {\bfseries 42} (1990) 2491}.

\bibitem{Kofman:1994rk}
L.~Kofman, A.D.~Linde and A.A.~Starobinsky, \emph{{Reheating after inflation}}, \href{https://doi.org/10.1103/PhysRevLett.73.3195}{\emph{Phys. Rev. Lett.} {\bfseries 73} (1994) 3195} [\href{https://arxiv.org/abs/hep-th/9405187}{{\ttfamily hep-th/9405187}}].

\bibitem{Saha:2020bis}
P.~Saha, S.~Anand and L.~Sriramkumar, \emph{{Accounting for the time evolution of the equation of state parameter during reheating}}, \href{https://doi.org/10.1103/PhysRevD.102.103511}{\emph{Phys. Rev. D} {\bfseries 102} (2020) 103511} [\href{https://arxiv.org/abs/2005.01874}{{\ttfamily 2005.01874}}].

\bibitem{Allahverdi:2010xz}
R.~Allahverdi, R.~Brandenberger, F.-Y.~Cyr-Racine and A.~Mazumdar, \emph{{Reheating in Inflationary Cosmology: Theory and Applications}}, \href{https://doi.org/10.1146/annurev.nucl.012809.104511}{\emph{Ann. Rev. Nucl. Part. Sci.} {\bfseries 60} (2010) 27} [\href{https://arxiv.org/abs/1001.2600}{{\ttfamily 1001.2600}}].

\bibitem{Amin:2014eta}
M.A.~Amin, M.P.~Hertzberg, D.I.~Kaiser and J.~Karouby, \emph{{Nonperturbative Dynamics Of Reheating After Inflation: A Review}}, \href{https://doi.org/10.1142/S0218271815300037}{\emph{Int. J. Mod. Phys. D} {\bfseries 24} (2014) 1530003} [\href{https://arxiv.org/abs/1410.3808}{{\ttfamily 1410.3808}}].

\bibitem{Drewes:2015coa}
M.~Drewes, \emph{{What can the CMB tell about the microphysics of cosmic reheating?}}, \href{https://doi.org/10.1088/1475-7516/2016/03/013}{\emph{JCAP} {\bfseries 03} (2016) 013} [\href{https://arxiv.org/abs/1511.03280}{{\ttfamily 1511.03280}}].

\bibitem{Repond:2016sol}
J.~Repond and J.~Rubio, \emph{{Combined Preheating on the lattice with applications to Higgs inflation}}, \href{https://doi.org/10.1088/1475-7516/2016/07/043}{\emph{JCAP} {\bfseries 07} (2016) 043} [\href{https://arxiv.org/abs/1604.08238}{{\ttfamily 1604.08238}}].

\bibitem{Aoki:2022dzd}
S.~Aoki, H.M.~Lee, A.G.~Menkara and K.~Yamashita, \emph{{Reheating and dark matter freeze-in in the Higgs-R$^{2}$ inflation model}}, \href{https://doi.org/10.1007/JHEP05(2022)121}{\emph{JHEP} {\bfseries 05} (2022) 121} [\href{https://arxiv.org/abs/2202.13063}{{\ttfamily 2202.13063}}].

\bibitem{Kallosh:1999jj}
R.~Kallosh, L.~Kofman, A.D.~Linde and A.~Van~Proeyen, \emph{{Gravitino production after inflation}}, \href{https://doi.org/10.1103/PhysRevD.61.103503}{\emph{Phys. Rev. D} {\bfseries 61} (2000) 103503} [\href{https://arxiv.org/abs/hep-th/9907124}{{\ttfamily hep-th/9907124}}].

\bibitem{PhysRevD.73.023501}
D.~Podolsky, G.N.~Felder, L.~Kofman and M.~Peloso, \emph{Equation of state and beginning of thermalization after preheating}, \href{https://doi.org/10.1103/PhysRevD.73.023501}{\emph{Phys. Rev. D} {\bfseries 73} (2006) 023501}.

\bibitem{Abbott:1984fp}
L.F.~Abbott and M.B.~Wise, \emph{{Constraints on Generalized Inflationary Cosmologies}}, \href{https://doi.org/10.1016/0550-3213(84)90329-8}{\emph{Nucl. Phys. B} {\bfseries 244} (1984) 541}.

\bibitem{Lucchin:1984yf}
F.~Lucchin and S.~Matarrese, \emph{{Power Law Inflation}}, \href{https://doi.org/10.1103/PhysRevD.32.1316}{\emph{Phys. Rev. D} {\bfseries 32} (1985) 1316}.

\bibitem{Ratra:1987rm}
B.~Ratra and P.J.E.~Peebles, \emph{{Cosmological Consequences of a Rolling Homogeneous Scalar Field}}, \href{https://doi.org/10.1103/PhysRevD.37.3406}{\emph{Phys. Rev. D} {\bfseries 37} (1988) 3406}.

\bibitem{Ferreira:1997hj}
P.G.~Ferreira and M.~Joyce, \emph{{Cosmology with a primordial scaling field}}, \href{https://doi.org/10.1103/PhysRevD.58.023503}{\emph{Phys. Rev. D} {\bfseries 58} (1998) 023503} [\href{https://arxiv.org/abs/astro-ph/9711102}{{\ttfamily astro-ph/9711102}}].

\bibitem{Enckell:2018hmo}
V.-M.~Enckell, K.~Enqvist, S.~Rasanen and L.-P.~Wahlman, \emph{{Inflation with $R^2$ term in the Palatini formalism}}, \href{https://doi.org/10.1088/1475-7516/2019/02/022}{\emph{JCAP} {\bfseries 02} (2019) 022} [\href{https://arxiv.org/abs/1810.05536}{{\ttfamily 1810.05536}}].

\bibitem{Aoki:2024jha}
S.~Aoki, A.~Ghoshal and A.~Strumia, \emph{{Cosmological collider non-Gaussianity from multiple scalars and $R^2$ gravity}},  \href{https://arxiv.org/abs/2408.07069}{{\ttfamily 2408.07069}}.

\bibitem{Anselmi:2020opi}
D.~Anselmi, \emph{{Quantum field theories of arbitrary-spin massive multiplets and Palatini quantum gravity}}, \href{https://doi.org/10.1007/JHEP07(2020)176}{\emph{JHEP} {\bfseries 07} (2020) 176} [\href{https://arxiv.org/abs/2006.01163}{{\ttfamily 2006.01163}}].

\bibitem{BICEP:2021xfz}
{\scshape BICEP, Keck} collaboration, \emph{{Improved Constraints on Primordial Gravitational Waves using Planck, WMAP, and BICEP/Keck Observations through the 2018 Observing Season}}, \href{https://doi.org/10.1103/PhysRevLett.127.151301}{\emph{Phys. Rev. Lett.} {\bfseries 127} (2021) 151301} [\href{https://arxiv.org/abs/2110.00483}{{\ttfamily 2110.00483}}].

\bibitem{Abazajian:2019eic}
K.~Abazajian et~al., \emph{{CMB-S4 Science Case, Reference Design, and Project Plan}},  \href{https://arxiv.org/abs/1907.04473}{{\ttfamily 1907.04473}}.

\bibitem{LiteBIRD:2022cnt}
{\scshape LiteBIRD} collaboration, \emph{{Probing Cosmic Inflation with the LiteBIRD Cosmic Microwave Background Polarization Survey}}, \href{https://doi.org/10.1093/ptep/ptac150}{\emph{PTEP} {\bfseries 2023} (2023) 042F01} [\href{https://arxiv.org/abs/2202.02773}{{\ttfamily 2202.02773}}].

\bibitem{Antoniadis:2018yfq}
I.~Antoniadis, A.~Karam, A.~Lykkas, T.~Pappas and K.~Tamvakis, \emph{{Rescuing Quartic and Natural Inflation in the Palatini Formalism}}, \href{https://doi.org/10.1088/1475-7516/2019/03/005}{\emph{JCAP} {\bfseries 03} (2019) 005} [\href{https://arxiv.org/abs/1812.00847}{{\ttfamily 1812.00847}}].

\bibitem{Tenkanen:2019jiq}
T.~Tenkanen, \emph{{Minimal Higgs inflation with an $R^2$ term in Palatini gravity}}, \href{https://doi.org/10.1103/PhysRevD.99.063528}{\emph{Phys. Rev. D} {\bfseries 99} (2019) 063528} [\href{https://arxiv.org/abs/1901.01794}{{\ttfamily 1901.01794}}].

\bibitem{Karam:2018mft}
A.~Karam, T.~Pappas and K.~Tamvakis, \emph{{Nonminimal Coleman--Weinberg Inflation with an $R^2$ term}}, \href{https://doi.org/10.1088/1475-7516/2019/02/006}{\emph{JCAP} {\bfseries 02} (2019) 006} [\href{https://arxiv.org/abs/1810.12884}{{\ttfamily 1810.12884}}].

\bibitem{Antoniadis:2018ywb}
I.~Antoniadis, A.~Karam, A.~Lykkas and K.~Tamvakis, \emph{{Palatini inflation in models with an $R^2$ term}}, \href{https://doi.org/10.1088/1475-7516/2018/11/028}{\emph{JCAP} {\bfseries 11} (2018) 028} [\href{https://arxiv.org/abs/1810.10418}{{\ttfamily 1810.10418}}].

\bibitem{Antoniadis:2019jnz}
I.~Antoniadis, A.~Karam, A.~Lykkas, T.~Pappas and K.~Tamvakis, \emph{{Single-field inflation in models with an $R^2$ term}}, \href{https://doi.org/10.22323/1.376.0073}{\emph{PoS} {\bfseries CORFU2019} (2020) 073} [\href{https://arxiv.org/abs/1912.12757}{{\ttfamily 1912.12757}}].

\bibitem{Gialamas:2020snr}
I.D.~Gialamas, A.~Karam and A.~Racioppi, \emph{{Dynamically induced Planck scale and inflation in the Palatini formulation}}, \href{https://doi.org/10.1088/1475-7516/2020/11/014}{\emph{JCAP} {\bfseries 11} (2020) 014} [\href{https://arxiv.org/abs/2006.09124}{{\ttfamily 2006.09124}}].

\bibitem{Karam:2021sno}
A.~Karam, E.~Tomberg and H.~Veerm\"ae, \emph{{Tachyonic preheating in Palatini $R^2$ inflation}}, \href{https://doi.org/10.1088/1475-7516/2021/06/023}{\emph{JCAP} {\bfseries 06} (2021) 023} [\href{https://arxiv.org/abs/2102.02712}{{\ttfamily 2102.02712}}].

\bibitem{Dimopoulos:2022rdp}
K.~Dimopoulos, A.~Karam, S.~S\'anchez~L\'opez and E.~Tomberg, \emph{{Palatini R$^{2}$ quintessential inflation}}, \href{https://doi.org/10.1088/1475-7516/2022/10/076}{\emph{JCAP} {\bfseries 10} (2022) 076} [\href{https://arxiv.org/abs/2206.14117}{{\ttfamily 2206.14117}}].

\bibitem{Ghoshal:2022qxk}
A.~Ghoshal, D.~Mukherjee and M.~Rinaldi, \emph{{Inflation and primordial gravitational waves in scale-invariant quadratic gravity with Higgs}}, \href{https://doi.org/10.1007/JHEP05(2023)023}{\emph{JHEP} {\bfseries 05} (2023) 023} [\href{https://arxiv.org/abs/2205.06475}{{\ttfamily 2205.06475}}].

\bibitem{Fujii:2003pa}
Y.~Fujii and K.~Maeda, \emph{{The scalar-tensor theory of gravitation}}, Cambridge Monographs on Mathematical Physics, Cambridge University Press (7, 2007), \href{https://doi.org/10.1017/CBO9780511535093}{10.1017/CBO9780511535093}.

\bibitem{abramowitz1968handbook}
M.~Abramowitz and I.A.~Stegun, \emph{Handbook of mathematical functions with formulas, graphs, and mathematical tables}, vol.~55, US Government printing office (1968).

\bibitem{Lima:2001lgd}
J.A.S.~Lima, R.~Silva and A.R.~Plastino, \emph{{Nonextensive thermostatistics and the H theorem}}, \href{https://doi.org/10.1103/PhysRevLett.86.2938}{\emph{Phys. Rev. Lett.} {\bfseries 86} (2001) 2938} [\href{https://arxiv.org/abs/cond-mat/0101030}{{\ttfamily cond-mat/0101030}}].

\bibitem{Lillepalu:2022knx}
H.G.~Lillepalu and A.~Racioppi, \emph{{Generalized hilltop inflation}}, \href{https://doi.org/10.1140/epjp/s13360-023-04512-1}{\emph{Eur. Phys. J. Plus} {\bfseries 138} (2023) 894} [\href{https://arxiv.org/abs/2211.02426}{{\ttfamily 2211.02426}}].

\bibitem{Lyth:2009zz}
D.H.~Lyth and A.R.~Liddle, \emph{{The primordial density perturbation: Cosmology, inflation and the origin of structure}} (2009).

\bibitem{Casadio:2005xv}
R.~Casadio, F.~Finelli, M.~Luzzi and G.~Venturi, \emph{{Higher order slow-roll predictions for inflation}}, \href{https://doi.org/10.1016/j.physletb.2005.08.056}{\emph{Phys. Lett. B} {\bfseries 625} (2005) 1} [\href{https://arxiv.org/abs/gr-qc/0506043}{{\ttfamily gr-qc/0506043}}].

\bibitem{Forconi:2021que}
M.~Forconi, W.~Giar\`e, E.~Di~Valentino and A.~Melchiorri, \emph{{Cosmological constraints on slow roll inflation: An update}}, \href{https://doi.org/10.1103/PhysRevD.104.103528}{\emph{Phys. Rev. D} {\bfseries 104} (2021) 103528} [\href{https://arxiv.org/abs/2110.01695}{{\ttfamily 2110.01695}}].

\bibitem{Lambiase:2023ryq}
G.~Lambiase, G.G.~Luciano and A.~Sheykhi, \emph{{Slow-roll inflation and growth of perturbations in Kaniadakis modification of Friedmann cosmology}}, \href{https://doi.org/10.1140/epjc/s10052-023-12112-7}{\emph{Eur. Phys. J. C} {\bfseries 83} (2023) 936} [\href{https://arxiv.org/abs/2307.04027}{{\ttfamily 2307.04027}}].

\bibitem{Planck:2018vyg}
{\scshape Planck} collaboration, \emph{{Planck 2018 results. VI. Cosmological parameters}}, \href{https://doi.org/10.1051/0004-6361/201833910}{\emph{Astron. Astrophys.} {\bfseries 641} (2020) A6} [\href{https://arxiv.org/abs/1807.06209}{{\ttfamily 1807.06209}}].

\bibitem{Kohri:2013mxa}
K.~Kohri, Y.~Oyama, T.~Sekiguchi and T.~Takahashi, \emph{{Precise Measurements of Primordial Power Spectrum with 21 cm Fluctuations}}, \href{https://doi.org/10.1088/1475-7516/2013/10/065}{\emph{JCAP} {\bfseries 10} (2013) 065} [\href{https://arxiv.org/abs/1303.1688}{{\ttfamily 1303.1688}}].

\bibitem{Basse:2014qqa}
T.~Basse, J.~Hamann, S.~Hannestad and Y.Y.Y.~Wong, \emph{{Getting leverage on inflation with a large photometric redshift survey}}, \href{https://doi.org/10.1088/1475-7516/2015/06/042}{\emph{JCAP} {\bfseries 06} (2015) 042} [\href{https://arxiv.org/abs/1409.3469}{{\ttfamily 1409.3469}}].

\bibitem{Munoz:2016owz}
J.B.~Mu\~noz, E.D.~Kovetz, A.~Raccanelli, M.~Kamionkowski and J.~Silk, \emph{{Towards a measurement of the spectral runnings}}, \href{https://doi.org/10.1088/1475-7516/2017/05/032}{\emph{JCAP} {\bfseries 05} (2017) 032} [\href{https://arxiv.org/abs/1611.05883}{{\ttfamily 1611.05883}}].

\bibitem{Linde:2011nh}
A.~Linde, M.~Noorbala and A.~Westphal, \emph{{Observational consequences of chaotic inflation with nonminimal coupling to gravity}}, \href{https://doi.org/10.1088/1475-7516/2011/03/013}{\emph{JCAP} {\bfseries 03} (2011) 013} [\href{https://arxiv.org/abs/1101.2652}{{\ttfamily 1101.2652}}].

\bibitem{Liddle:2003as}
A.R.~Liddle and S.M.~Leach, \emph{{How long before the end of inflation were observable perturbations produced?}}, \href{https://doi.org/10.1103/PhysRevD.68.103503}{\emph{Phys. Rev. D} {\bfseries 68} (2003) 103503} [\href{https://arxiv.org/abs/astro-ph/0305263}{{\ttfamily astro-ph/0305263}}].

\bibitem{Baumann:2022mni}
D.~Baumann, \emph{{Cosmology}}, Cambridge University Press (7, 2022), \href{https://doi.org/10.1017/9781108937092}{10.1017/9781108937092}.

\bibitem{Mukhanov:2005sc}
V.~Mukhanov, \emph{{Physical Foundations of Cosmology}}, Cambridge University Press, Oxford (2005), \href{https://doi.org/10.1017/CBO9780511790553}{10.1017/CBO9780511790553}.

\bibitem{Baumann:2009ds}
D.~Baumann, \emph{{Inflation}},  in \emph{{Theoretical Advanced Study Institute in Elementary Particle Physics}: {Physics of the Large and the Small}}, pp.~523--686, 2011, \href{https://doi.org/10.1142/9789814327183_0010}{DOI} [\href{https://arxiv.org/abs/0907.5424}{{\ttfamily 0907.5424}}].

\bibitem{baumann2014cosmology}
D.~Baumann, \emph{Cosmology, part iii mathematical tripos}, {\emph{University lecture notes} {\bfseries 56} (2014) }.

\bibitem{Bostan:2018evz}
N.~Bostan, O.~G\"ulery\"uz and V.N.~\c{S}eno\u{g}uz, \emph{{Inflationary predictions of double-well, Coleman-Weinberg, and hilltop potentials with non-minimal coupling}}, \href{https://doi.org/10.1088/1475-7516/2018/05/046}{\emph{JCAP} {\bfseries 05} (2018) 046} [\href{https://arxiv.org/abs/1802.04160}{{\ttfamily 1802.04160}}].

\bibitem{Bostan:2024fyz}
N.~Bostan, \emph{{Reheating constraints to Palatini Coleman-Weinberg inflation}}, \href{https://doi.org/10.55730/1300-0101.2754}{\emph{Turk. J. Phys.} {\bfseries 48} (2024) 28}.

\bibitem{Maths}
N.N.~Lebedev, \emph{Special Functions and Their Applications}, Dover Publications, Inc. (1972).

\bibitem{Press2007}
W.H.~Press, S.A.~Teukolsky, W.T.~Vetterling and B.P.~Flannery, \emph{Numerical Recipes 3rd Edition: The Art of Scientific Computing}, Cambridge University Press, 3~ed. (2007).

\end{thebibliography}



\providecommand{\href}[2]{#2}\begingroup\raggedright\endgroup

\end{document}